\documentclass[prd, twocolumn, superscriptaddress]{revtex4-2}
\usepackage{amsmath}
\usepackage{amssymb}
\usepackage{dsfont}
\usepackage{graphicx}
\allowdisplaybreaks[1]
\usepackage{xcolor}

\begin{document}
\title{Observing single-particles beyond the Rindler horizon}

\author{Riccardo Falcone}
\affiliation{Department of Physics, University of Sapienza, Piazzale Aldo Moro 5, 00185 Rome, Italy}

\author{Claudio Conti}
\affiliation{Department of Physics, University of Sapienza, Piazzale Aldo Moro 5, 00185 Rome, Italy}
\affiliation{Institute for Complex Systems (ISC-CNR), Department of Physics, University Sapienza, Piazzale Aldo Moro 2, 00185, Rome, Italy}
\affiliation{Research Center Enrico Fermi, Via Panisperna 89a, 00184 Rome, Italy}

\begin{abstract}
We show that Minkowski single-particle states localized beyond the horizon modify the Unruh thermal distribution in an accelerated frame. This means that, contrary to classical predictions, accelerated observers can reveal particles emitted beyond the horizon. The method we adopt is based on deriving the explicit Wigner characteristic function for the complete description of the quantum field in the non-inertial frame and can be generalized to general states.
\end{abstract}

\maketitle

In classical physics, an observer in an accelerated frame cannot detect signals emitted beyond an event horizon. One can argue if this is also true in quantum physics. Starting from pioneering investigations \cite{Unruh, PhysRevD.29.1047}, many authors have studied how accelerated observers can register inertial vacuum states through thermal particles detection. Here, we ask if inertial single-particle states localized beyond the horizon can be revealed by monitoring variations of the particle distribution from the thermal background. To answer this question, we adopt quantum field theory in curved space-time \cite{wald}, developed in the last decades with groundbreaking results as the Hawking \cite{1103899181} and the Unruh effect \cite{PhysRevD.7.2850, 1975, Unruh}.

In the Unruh effect, a thermal state replaces the vacuum when the observer is accelerated, as an outcome of the fact that quantum states are reference-frame dependent \cite{PhysRevD.7.2850}. Beyond the vacuum, various authors considered more general states focusing on entangled systems (see \cite{PhysRevA.87.012306, PhysRevA.87.052326, PhysRevD.95.076004} and references therein). Indeed, entanglement is significant because of the quantum correlation between two regions of space-time – denoted as the Rindler left and right wedge – and the need to trace over one of the two wedges to predict observable quantities in accelerate frames.

A way to describe the transition from Minkowski to Rindler frames can be made using Wigner
distributions \cite{2020}. Recently, Ben-Benjamin, Scully, and Unruh have reported \cite{2020} the Wigner distribution for the Minkowski vacuum state in the right wedge and the Minkowski number states in both the right and the left wedges. However, to the best of our knowledge, the explicit expression for Minkowski number states in the right wedge -- tracing out the left wedge -- is still missing.

In this Letter, we compute the characteristic function \cite{Barnett2002} of single-particle states in accelerated frames. From the characteristic function, we derive the probability of finding a Rindler particle when a Minkowski particle is emitted. We show that there is a finite probability of detecting a Rindler particle as a perturbation to the Unruh thermal background, even when the Minkowski particle is localized beyond the horizon.

\begin{figure}[h]
\includegraphics[]{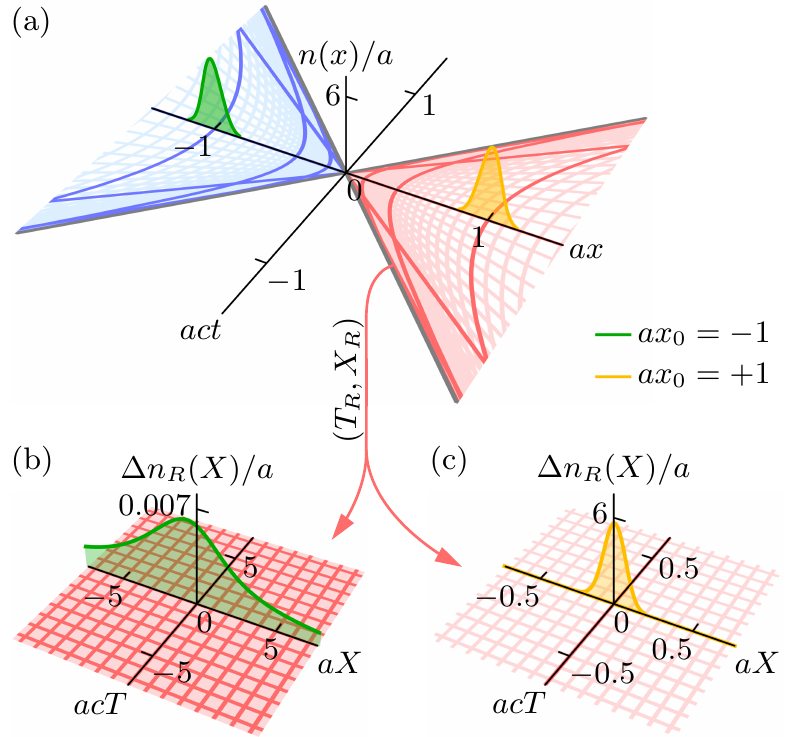}
\caption{Representation of the coordinate transformation $(t,x) \mapsto (T,X) = (T_R(t,x),X_R(t,x))$ and the probability density function transformation $n(x) \mapsto \Delta n_R(X)$ for localized wave-packets from the inertial to the accelerated frame. The yellow and the green lines are associated to single-particle states with Gaussian wave-function $\psi(x)$ defined by Eq. (\ref{Gaussian_psi_x}) with $a \sigma = 1$ and $ax_0 = \pm 1$. The Rindler left and right wedges are shown in panel (a) through constant $T$ and $X$ lines. The two wedges are delimited by the Rindler horizons (gray). The profile of the probability density $n(x)$ -- defined by Eq. (\ref{n}) -- for the two wave-packets has been drawn. In panels (b) and (c), we show the constant $T$ and $X$ lines in the accelerated frame and the profile of $\Delta n_R(X)$, defined by Eq. (\ref{Delta_n}). The value of $\Delta n_R(X)$ gives the variation of probability density to find a Rindler particle in $X$ with respect to the Minkowski vacuum. In panel (b), $\Delta n_R(X)$ is not-negligible, even if $\psi(x)$ is localized in the left wedge. On the other hand, in panel (c), $\Delta n_R(X)$ is larger and narrower for $a x_0=+1$, as we expect from wave-functions localized within the right wedge.}\label{rindler_figure}
\end{figure}

By following the original works of Fulling, Davis and Unruh \cite{PhysRevD.7.2850, 1975, Unruh}, we consider a (1-1)-dimensional flat space-time and the coordinate transformation $(t_R,x_R)$ from an accelerated frame $(T,X)$ to an inertial frame $(t,x)=(t_R(T,X),x_R(T,X))$: $a c t_R(T,X) = e^{aX} \sinh a c T$ and $ a x_R(T,X) = e^{aX} \cosh a c T $, where $a c^2$ is the acceleration -- which is conventionally taken positive -- and $c$ the speed of light. Such transformation covers only the right Rindler wedge. On the other hand, it is possible to cover the left Rindler wedge through the transformation $t = t_L(T,X)$ and $x = x_L(T,X)$ with $t_L(T,X)$ and $x_L(T,X)$ being identical to $t_R(T,X)$ and $x_R(T,X)$ but with opposite acceleration. In the notation we have adopted, the subscript $L$ ($R$) refers to the left (right) wedge. It is possible to see a visual representation of the Minkowski and Rindler coordinates in Fig. \ref{rindler_figure}, where $(T_L,X_L)$ and $(T_R,X_R)$ are taken as the inverse transformations of $(t_L,x_L)$ and $(t_R,x_R)$. However, since we are not interested in temporal evolutions of states, for the rest of the paper we will just refer to $x_{L,R}(X)$ and $X_{L,R}(x)$ as, respectively $x_{L,R}(0,X)$ and $X_{L,R}(0,x)$.

By following again the original works of Fulling, Davis and Unruh \cite{PhysRevD.7.2850, 1975, Unruh}, we consider a massless free scalar field $\hat{\phi}(t,x)$. We name $\hat{a}(k)$ the annihilation operator for the Minkowski mode with momentum $k$, while $\hat{b}_L(K)$ ($\hat{b}_R(K)$) the annihilation operators for the left (right) Rindler mode with momentum $K$. 

The Unruh effect can be obtained by representing the Minkowski vacuum state $| 0_M \rangle$ -- defined by $ \hat{a}(k)| 0_M \rangle=0$ for any $k \in \mathbb{R}$ -- in the representation space of the $ \hat{b}_{L,R}(K)$-algebra. This leads to the following state \cite{Unruh}
\begin{align}  \label{vacuum_state}
|0_M\rangle \propto & \exp \left( \int_{-\infty}^{+\infty} dK \exp \left( - \frac{\beta}{2} |K| \right) \right. \nonumber \\
& \times \left. \hat{b}^\dagger_L(K) \hat{b}^\dagger_R(K) \right) | 0_L, 0_R \rangle,
\end{align}
with $\beta = 2 \pi / a$ and $| 0_{L,R} \rangle$ defined by $\hat{b}_{L,R}(K)| 0_{L,R} \rangle=0$. The final expression for the Minkowski vacuum state in the right Rindler frame can be obtained by performing a partial trace over the left wedge, which leads to a thermal state $\hat{\rho}_0$ with temperature $T_0 = (k_B \beta)^{-1}$, where $k_B$ is the Boltzmann constant.

Analogously to $| 0_M \rangle$, any Minkowski single-particle state $| \psi \rangle$ can be represented in the right wedge through a representative in the $\hat{b}_{L,R}(K)$-algebra and by performing a partial trace over the left wedge $\hat{\rho} = \text{Tr}_L |\psi\rangle \langle \psi |$. Here, $| \psi \rangle$ is defined through a normalized wave function $\psi(x)$ such that
\begin{equation} \label{one_state}
|\psi\rangle = \int_{-\infty}^{+\infty} dx \psi(x) \hat{\tilde{a}}^\dagger (x) |0_M\rangle,
\end{equation}
where $\hat{\tilde{a}}^\dagger (x) = \int_{-\infty}^{+\infty} dk e^{-i k x} \hat{a}^\dagger (k) /\sqrt{2 \pi}$ is the creation operator for a particle in position $x$.

Eq. (\ref{one_state}) can be put into the following form
\begin{equation} \label{one_state_2}
|\psi\rangle = \int_{-\infty}^{+\infty} dK \left[ \tilde{\psi}_-(K) \hat{b}_R(K)+ \tilde{\psi}_+(K) \hat{b}^\dagger_R(K) \right]   |0_M\rangle,
\end{equation}
with 
\begin{subequations} \label{one_state_2_}
\begin{align} \label{psi_tilde_2}
\tilde{\psi}_\pm(K) = & \frac{e^{-\theta(\pm 1)\beta |K|}}{n_0(K)} \int_{-\infty}^{+\infty} dX \frac{e^{ \mp i K X }}{\sqrt{2 \pi}} \left\lbrace \psi_R (X) \left[ \theta(\pm 1)  \right.\right.  \nonumber \\
& \left. \left.  + \tilde{f}_{R \pm}\left( \mp \frac{K}{a} \right) \right] + \psi_L (-X) \tilde{f}_{L \pm}\left( \mp \frac{K}{a} \right)  \right\rbrace,
\end{align}
\begin{align} \label{f_tilde_LR}
\tilde{f}_{L,R \pm}(\kappa) = & - \theta(s_{L,R}) \theta(\pm 1) + \frac{1}{2 \pi}\sqrt{|\kappa|} \Gamma (i\kappa) \Gamma \left(\frac{1}{2} - i\kappa \right) \nonumber \\
& \times \exp \left( \pm \theta(s_{L,R}) \pi |\kappa| \pm  i s_{L,R} \text{sign}(\kappa) \frac{\pi}{4} \right),
\end{align}
\begin{equation} \label{psi_LR}
\psi_{L,R}(X) = \sqrt{a |x_{L,R}(X)|} \psi(x_{L,R}(X)),
\end{equation}
\begin{equation}
s_L=-1, \quad s_R=1, \quad n_0(K) = (e^{\beta|K|}-1)^{-1}.
\end{equation}
\end{subequations}

A proof for Eq. (\ref{one_state_2}) is given in the Supplemental Material (SM). The key element for such proof is provided by the following identity
\begin{equation} \label{vacuum_state_property}
\hat{b}^\dagger_{L,R}(K) |0_M\rangle = \exp \left( \frac{\beta}{2} |K| \right) \hat{b}_{R,L}(K) |0_M\rangle ,
\end{equation}
which holds for any $K \in \mathbb{R}$. Eq. (\ref{vacuum_state_property}) states that the creation of a Rindler particle in the left (right) wedge over the Minkowski vacuum background is equivalent to the destruction of a Rindler particle in the right (left) wedge, up to an $\exp(\beta|K|/2)$ factor. Thanks to Eq. (\ref{vacuum_state_property}), we can give the following interpretation to the functions $\tilde{\psi}_\pm(K)$ that appear in Eq. (\ref{one_state_2}). $\tilde{\psi}_+(K)$ ($\tilde{\psi}_-(K)$) can be seen as the wave-function of a Rindler particle created (destroyed) over the Minkowski vacuum background in the right wedge, or -- up to an $\exp(\beta|K|/2)$ factor -- as a Rindler particle destroyed (created) in the left wedge.

$\psi_R(X)$, on the other hand, can be interpreted as a transformed version of the wave-function $\psi(x)$ in terms of the infinitesimal probability function $n(x) dx = |\psi(x)|^2 dx$. Indeed, from Eq. (\ref{psi_LR}), it is possible to notice that for $x>0$, $\left| \psi(x) \right|^2 dx$ is equivalent to $\left| \psi_R(X) \right|^2 dX$, up to the coordinate transformation $x \mapsto X=X_R(x)$.

By taking the partial trace of $|\psi\rangle \langle \psi|$ over the left wedge, eq. (\ref{one_state_2}) results in the following expression for the transformed single-particle state
\begin{align} \label{rho_2}
\hat{\rho} = & \int_{-\infty}^{+\infty} dK \left[ \tilde{\psi}_-(K) \hat{b}_R(K)+ \tilde{\psi}_+(K) \hat{b}^\dagger_R(K) \right]  \hat{\rho}_0 \nonumber \\
& \times \int_{-\infty}^{+\infty} dK' \left[ \tilde{\psi}^*_-(K') \hat{b}^\dagger_R(K') +  \tilde{\psi}^*_+(K') \hat{b}_R(K') \right].
\end{align}

As we have mentioned before, an alternative representation for the state $\hat{\rho}$ can be provided through the following characteristic function \cite{Barnett2002}
\begin{align} \label{chi}
\chi[\xi,\xi^*] = & \text{Tr} \left[ \hat{\rho} \exp \left( \int_{-\infty}^{+\infty} dK  \xi(K) \hat{b}^\dagger_R(K)\right) \right. \nonumber \\
& \left. \times \exp \left(- \int_{-\infty}^{+\infty} dK \xi^*(K) \hat{b}_R(K) \right) \right].
\end{align}
Thanks to Eq. (\ref{rho_2}), we can write $\chi[\xi,\xi^*]$ in terms of functional derivatives of the characteristic function for the thermal state $\chi_0[\xi,\xi^*]$ (SM):
\begin{align} \label{characteristic_function_one_particle_final_2}
\chi[\xi,\xi^*]  = & \left\lbrace 1 - \left| \int_{-\infty}^{+\infty} dK  n_0(K) \left[\tilde{\psi}_-(K) \xi(K) \right. \right. \right. \nonumber \\
& \left. \left. \left. - e^{\beta |K|} \tilde{\psi}_+(K) \xi^*(K) \right] \right|^2 \right\rbrace \chi_0[\xi,\xi^*].
\end{align}
Finally, the explicit expression for $\chi[\xi,\xi^*]$ can be obtained from Eq. (\ref{characteristic_function_one_particle_final_2}) supplemented with Eqs. (\ref{one_state_2_}) and the already-known expression for $\chi_0[\xi,\xi^*]$ \cite{Barnett2002}:
\begin{equation} \label{chi_0_2}
\chi_0[\xi,\xi^*] = \exp \left( - \int_{-\infty}^{+\infty}dK n_0(K) |\xi(K)|^2 \right).
\end{equation}

Functional derivatives of the characteristic function $\chi[\xi,\xi^*]$ allow us to extract different mean values of $\hat{\rho}$ \cite{Barnett2002}. In this way, we compute the probability density function of $\hat{\rho}$, defined as
\begin{align} \label{n_postion}
\left\langle \hat{n}_R(X) \right\rangle_{\hat{\rho}} = & \int_{-\infty}^{+\infty} dK \frac{e^{-i K X}}{\sqrt{2\pi}} \int_{-\infty}^{+\infty} dK' \frac{e^{i K' X}}{\sqrt{2\pi}} \nonumber \\
& \left. \times \frac{\delta}{\delta \xi(K)} \left(-\frac{\delta}{\delta \xi^*(K')} \right) \chi[\xi,\xi^*] \right|_{\xi=0},
\end{align}
where $\hat{n}_R(X) = \hat{\tilde{b}}_R^\dagger (X) \hat{\tilde{b}}_R (X)$ is the particle density operator and $\hat{\tilde{b}}_R (X) = \int_{-\infty}^{+\infty} dK e^{i K X}  \hat{b}_R (K)/\sqrt{2\pi}$ is the annihilation operator in $X$. Eq. (\ref{n_postion}) results into the following equation (SM)
\begin{equation}\label{n_postion_2}
\Delta n_R(X)  = n_+(X) + n_-(X),
\end{equation}
with
\begin{subequations}
\begin{equation} \label{Delta_n}
\Delta n_R(X) = \left\langle \hat{n}_R(X) \right\rangle_{\hat{\rho}} - \left\langle \hat{n}_R(X) \right\rangle_{\hat{\rho}_0},
\end{equation}
\begin{equation}\label{n_pm}
n_\pm (X) = \left| \int_{-\infty}^{+\infty} dK\frac{e^{\pm i K X}}{\sqrt{2 \pi}} n_0(K) e^{\theta(\pm 1) \beta |K|} \tilde{\psi}_\pm(K) \right|^2.
\end{equation}
\end{subequations}

$\Delta n_R(X)$ -- defined by Eq. (\ref{Delta_n}) -- represents the difference in the probability density function between the Minkowski single-particle and the Minkowski vacuum state in terms of Rindler particles. An accelerated observer measuring a non-vanishing $\Delta n_R(X)$ can infer the presence of a Minkowski particle. Fig. \ref{rindler_figure} shows $\Delta n_R(X)$ for Gaussian wave-functions in comparison with the probability density function in the Minkowski space-time, defined as
\begin{equation} \label{n}
n(x) = \left\langle \hat{\tilde{a}}^\dagger (x) \hat{\tilde{a}}(x) \right\rangle_{| \psi \rangle \langle \psi |}.
\end{equation}

$n_\pm (X)$ derive from $\hat{\rho}$ of Eq. (\ref{rho_2}) through the contribution of, respectively, $\tilde{\psi}_\pm(K)$. Therefore, they are associated to the Rindler particles respectively created and destroyed over the Minkowski vacuum background in the right wedge. Their explicit form with respect to $\psi_{L,R}(X)$ reads (SM)
\begin{equation}\label{n_pm_2}
n_\pm (X) =  | \theta(\pm 1) \psi_R(X) + \psi_{R \pm}(X) + \psi_{L \pm}(X) |^2 ,
\end{equation}
with
\begin{subequations}
\begin{equation} \label{psi_LR_pm}
\psi_{L,R \pm}(X) = \int_{-\infty}^{+\infty} d\xi \psi_{L,R}\left(s_{L,R}\frac{\xi}{a} \right) f_{L,R \pm}( \xi - aX ),
\end{equation}
\begin{equation}\label{f_LR}
f_{L,R \pm}( \xi ) = \int_{-\infty}^{+\infty} d\kappa \frac{e^{i \kappa \xi}}{2 \pi} \tilde{f}_{L,R \pm}(\kappa).
\end{equation}
\end{subequations}

\begin{figure}[h]
\includegraphics[]{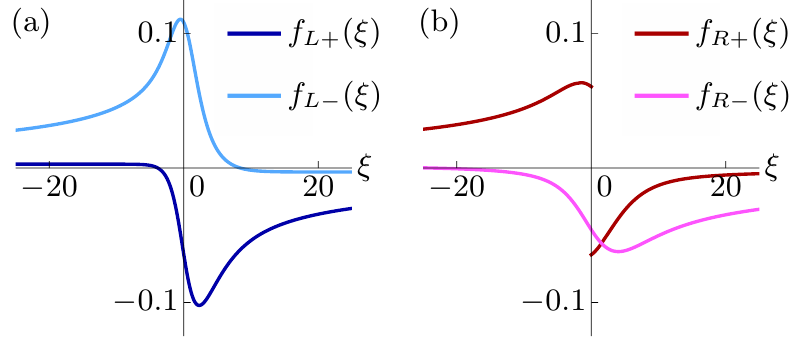}
\caption{Profile of $f_{L,R \pm}( \xi )$ defined by Eq. (\ref{f_LR}) which have been numerically derived through a Fourier transform of $\tilde{f}_{L,R \pm}( \xi )$ defined by Eq. (\ref{f_tilde_LR})}\label{f_figure}
\end{figure}

$\psi_R(X)$ and $\psi_{L,R +}(X)$ of Eq. (\ref{n_pm_2}) play the role of superposed wave-functions which result in a probability density function $n_+ (X)$ describing Rindler particles created over $|0_M\rangle$ in the right wedge. $\psi_{L,R -}(X)$, on the other hand, refer to Rindler particles destroyed in the right wedge. Within the decomposition of the wave-functions $\psi_R(X) + \psi_{L +}(X) + \psi_{R +}(X)$ and  $\psi_{L -}(X) + \psi_{R -}(X)$, $\psi_{L\pm}(X)$ derive from the left-wedge part of $\psi(x)$ -- i.e. $\psi(x)$ for $x<0$ -- while $\psi_R(X)$ and $\psi_{R \pm}(X)$ from the right-wedge part of $\psi(x)$.

The presence of $\psi_{L \pm}(X)$ in Eq. (\ref{n_pm_2}) implies that values of the wave-function beyond the horizon give non-vanishing contributions to $\left\langle \hat{n}_R(X) \right\rangle_{\hat{\rho}}$. Even a state with small values of $|\psi(x)|$ for $x<0$ can still be detected in the right Rindler wedge. We remark that this effect is not due to the right tail of the wave function, since the corresponding contribution is exponentially smaller than the leading one, as detailed below with a specific example.

We can argue that the result may change if we use a Lorentz-invariant normalization for $\psi(x)$ \cite{Localized}, since left-wedge values of the wave-function are normalization-dependent. Nevertheless, we have verified that left-wedge values of $\psi(x)$ appear in $\Delta n_R(X)$ even when we use a Lorentz-invariant normalization (SM).

$f_{L,R \pm}( \xi )$, shown in Fig. \ref{f_figure}, are localized around $\xi = 0$. This means that $\Delta n_R(X)$ receives most contributions from $\psi_L (X')$ and $\psi_R (X')$ from $X' \approx -X$ and $X' \approx X$, respectively, and within a region $\Delta X' \sim a^{-1}$. In the case of Fig. \ref{f_figure}a, this implies that most of the contributions for $\psi_{L \pm}(X)$ come from $\psi_L (X')$ when $x_L(X') = - x_R(X)$, or, equivalently, from $\psi (-x_R(X))$. Moreover, wave-functions localized in the left wedge -- i.e. with small values of $|\psi(x)|$ for $x>0$ -- are characterized by a $\Delta n_R(X)$ whose main contributions come from $\psi_{L \pm}(X)$, since $\psi_R(X)$ is defined by right-wedge values of $\psi(x)$. This means that $\Delta n_R(X) \approx \left| \psi_{L+}(X) \right| ^2 + \left| \psi_{L-}(X) \right|^2 $ and that most of the contributions for $\Delta n_R(X) $ come from $\psi (x)$, with $x$ as the specular point of $X$ in the Minkowski space-time with respect to the horizon -- i.e. $x = - x_R(X)$.

\begin{figure}[h]
\includegraphics[]{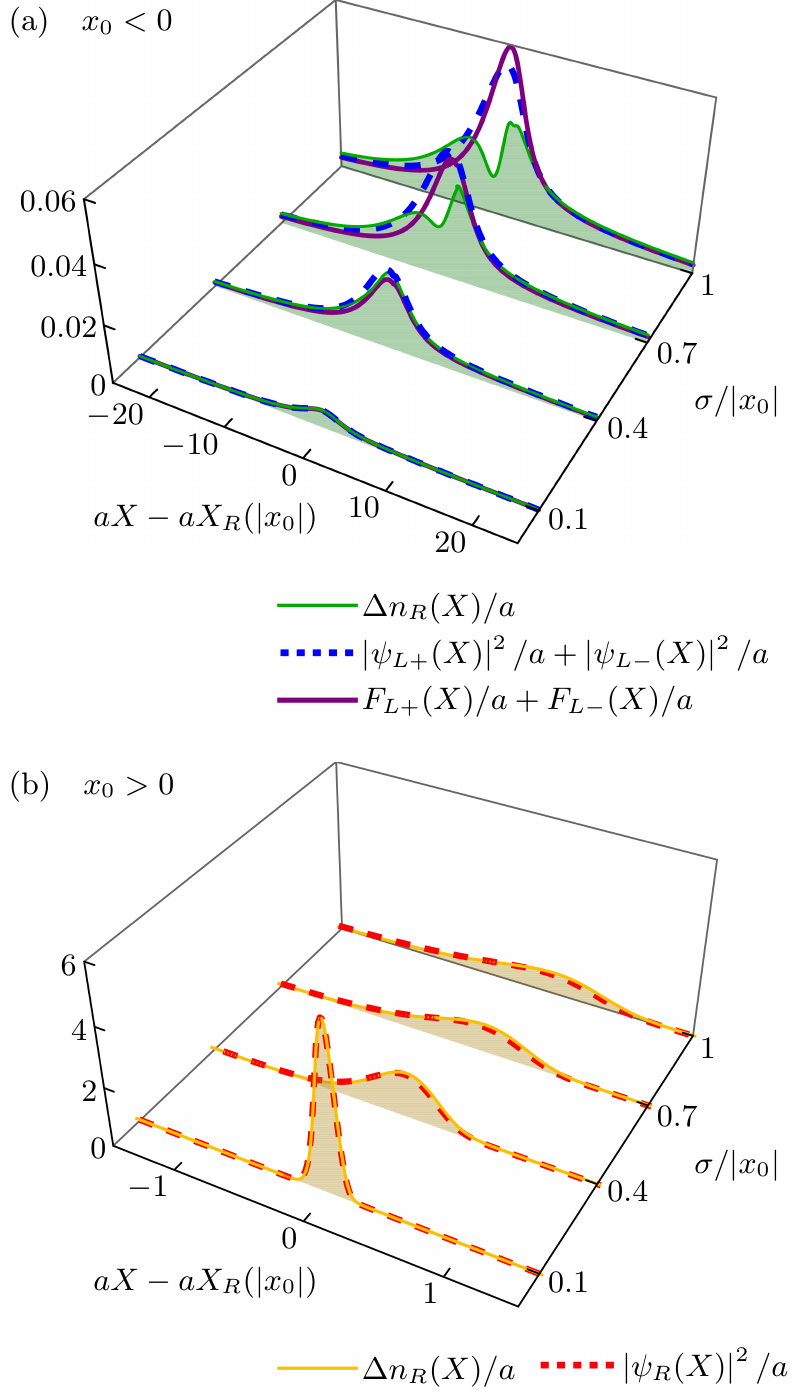}
\caption{Profile of $\Delta n_R (X)$ defined by Eq. (\ref{Delta_n}) for different Gaussian single-particle states (\ref{Gaussian_psi_x}). In panel (a), the configurations are defined by $x_0<0$ and different values of $\sigma/|x_0|$. $\left| \psi_{L+}(X) \right| ^2 + \left| \psi_{L-}(X) \right|^2$ is the dominant contribution of Eq. (\ref{n_postion_2}) when $\sigma/|x_0| \rightarrow 0$ and $x_0<0$. We also show the function $F_{L+}(X)+F_{L-}(X)$ defined in Eq.(\ref{F_LRpm}). In panel (b), the profile of $\Delta n_R (X)$ and $\left| \psi_R(X) \right|^2$ for configurations with $x_0>0$ is shown. In this case, the dominant contribution to $\Delta n_R (X)$ is $\left| \psi_R(X) \right|^2$, which, in the particular case of Gaussian wave-functions, has Eq. (\ref{components_limits_psi_R}) as distributional limit.}\label{Delta_n_R_figure}
\end{figure}

It is also possible to notice from Fig. \ref{f_figure}b that $ |f_{R \pm}(\xi)| \ll 1$. Therefore, if $\psi(x)$ is localized in a region $\mathcal{R}$ -- i.e. $|\psi(x)|$ is small outside a finite region $\mathcal{R}$ -- and if $\mathcal{R}$ is in the right wedge and with a width $\Delta x \ll a^{-1}$, then $\psi_{R \pm}(X)$ are expected to be negligible with respect to $\psi_R(X)$. The same happens for states with $\Delta x \sim a^{-1}$ but with $\mathcal{R}$ far away from the origin with respect to $a$ -- i.e. $x \gg a^{-1}$ for any $x \in \mathcal{R}$. This last result can be motivated by the fact that the transformed region $x_R (\mathcal{R})$ becomes way smaller than $a^{-1}$ when $x \gg a^{-1}$ for any $x \in \mathcal{R}$ and, therefore, $\psi_R(X)$ is non-negligible within a region way smaller than $a^{-1}$. In summary, $|\psi_{R \pm}(X)| \ll |\psi_R(X)|$ for any $X$ and for wave-functions well-localized in the right wedge -- i.e. when $\Delta x \ll a^{-1}$ and $x \gtrsim a^{-1}$ for any $x \in \mathcal{R}$ or when $x \gg a^{-1}$ for any $x \in \mathcal{R}$. Moreover, for such states, $\psi_{L\pm}(X)$ are negligible and therefore $\Delta n_R(X) \approx \left| \psi_R(X) \right|^2 $. In other words, $\psi_R(X)$ acts like a probability amplitude for wave-functions well-localized in the right wedge and it appears as the dominant term in Eq. (\ref{n_postion_2}), with $\psi_{L,R\pm}(X)$ as small corrective terms.

We remark that the wave-function can still have infinite tails. To give a quantitative example, we consider a normalized Gaussian wave-functions, whose localization degree is given by the variance $\sigma$: 
\begin{equation} \label{Gaussian_psi_x}
\psi(x) = \frac{1}{\sqrt[4]{\pi}\sqrt{\sigma}} \exp \left( - \frac{(x-x_0)^2}{2 \sigma^2} \right).
\end{equation}

We are interested in the limit $a \sigma \rightarrow 0$ for fixed $x_0 \neq 0$ and the limit $a x_0 \rightarrow \pm \infty$ for fixed $\sigma$, which correspond to the case of well-localized states in the left or right wedge. It is possible to prove that for Gaussian wave-functions the limit $a \sigma \rightarrow 0$ is equivalent to $a |x_0| \rightarrow \infty$ up to a translation of $\Delta n_R(X)$ with respect to $X$ (SM). More specifically, it is possible to prove that when $x_0 \neq 0$, any transformation $x_0 \mapsto \alpha x_0$ with $\alpha > 0$ acting on $\psi_R(X)$ is equivalent to $\sigma \mapsto \sigma / \alpha$, $a X \mapsto a X - \ln \alpha$. This also applies to $\psi_{L,R \pm}(X)$ and $\Delta n_R(X)$. Given the invariance under the transformation $x_0 \mapsto x_0/\alpha$, $\sigma \mapsto \sigma / \alpha$, $a X \mapsto a X - \ln \alpha$ for any $\alpha>0$, the functions $\psi_R(X)$, $\psi_{L,R \pm}(X)$ and $\Delta n_R(X)$ can be put in a form depending of $\text{sign}(x_0)$, $\sigma/|x_0|$ and $X-X_R(|x_0|)$ instead of $\sigma$, $x_0$ and $X$. This feature is adopted in Fig. \ref{Delta_n_R_figure}, where we show $\Delta n_R(X)$ for different $\sigma/x_0$.

The limit of well-localized wave-functions is identified with $\sigma/|x_0| \rightarrow 0$. In Fig. \ref{Delta_n_R_figure}, we show how Gaussian wave-functions give the same results expected for the general case. Specifically, Fig. \ref{Delta_n_R_figure}a shows that for wave-packets well-localized in the left wedge, $\Delta n_R(X) \approx \left| \psi_{L+}(X) \right| ^2 + \left| \psi_{L-}(X) \right|^2 $. Fig. \ref{Delta_n_R_figure}b shows that $\Delta n_R(X) \approx \left| \psi_R(X) \right|^2 $ when $\sigma/|x_0| \rightarrow 0$ and $x_0 > 0$. This result can be proven analytically (SM): 
\begin{widetext}

\begin{subequations}\label{components_limits}
\begin{equation}\label{components_limits_psi_R}
| \psi_R(X) |^2  = \begin{cases}
\delta(X-X_R(x_0)) + \mathcal{O} \left( \frac{|x_0|}{\sigma} \exp \left( -  \frac{|x_0|^2}{2 \sigma^2} \right) \right) & \text{if } x_0>0 \\
\mathcal{O} \left( \frac{|x_0|}{\sigma} \exp \left( -  \frac{x_0^2}{2 \sigma^2} \right) \right) & \text{if } x_0<0
\end{cases}
\end{equation}
\begin{equation}\label{components_limits_psi_LRpm}
|\psi_{L,R \pm}(X)|^2 = \begin{cases}
F_{L,R \pm}(X) + o\left(\frac{\sigma}{|x_0|}\right) & \text{if } s_{L,R} x_0 >0 \\
o \left( \frac{\sigma}{|x_0|} \exp \left( -  \frac{x_0^2}{2 \sigma^2} \right) \right) & \text{if } s_{L,R} x_0<0
\end{cases}
\end{equation}
\end{subequations}

\end{widetext}
with
\begin{equation}\label{F_LRpm}
F_{L,R \pm} (X) = \frac{\sigma}{|x_0|}  2 \sqrt{\pi} a f^2_{L,R +}(a X_R(|x_0|)- a X).
\end{equation}

From Eqs. (\ref{components_limits}) we obtain the explicit limit $\sigma/|x_0| \rightarrow 0$ of $\Delta n_R(X)$ for Gaussian wave-functions. When $x_0<0$, $\Delta n_R(X) \rightarrow 0$ with leading term $F_{L+}(X) + F_{L-}(X)$ which is proportional to $ \sigma / |x_0| $. When the degree of localization of the particle increases, the probability of detection in the right wedge decreases. Nevertheless, if $\sigma \ll |x_0|$ but $\sigma \neq 0$, the profile of $\Delta n_R (X)$ is approximately $F_{L+}(X) + F_{L-}(X)$, as in Fig. \ref{Delta_n_R_figure}a. Accelerated observers can still see a difference with respect to the vacuum state, even when the particle is localized beyond the horizon. The result does not depend on the presence of a tail in the right Rindler wedge, since most of the contributions for $\Delta n_R (X)$ come from values of $\psi(x)$ beyond the horizon. Indeed, $\psi_R(X)$ and $\psi_{R \pm}(X)$ are vanishing with exponential orders, while $\psi_{L \pm}(X)$ are linear in $ \sigma / |x_0| $.

The peak of $F_{L+}(X) + F_{L-}(X)$ in $X=X_R(|x_0|)$ results in a maximum probability to find the particle in $X=X_R(-x_0)$. In the Minkowski space-time, such point correspond to the specular counterpart of $x_0$ with respect to the horizon: $x_R(X) = - x_0$.

When $x_0>0$, the distributional limit of $\Delta n_R(X)$ is $\delta(X-X_R(x_0))$, as in Fig. \ref{Delta_n_R_figure}b. The single-particle appears perfectly localized in both inertial and accelerated frame at the same position (up to the coordinate transformation).

In conclusion, we have provided a complete description for single-particle states in accelerated frames $\hat{\rho}$ through their characteristic functions $\chi[\xi,\xi^*]$. By the derivatives of $\chi[\xi,\xi^*]$, we obtain original expressions for the right-wedge density function $\left\langle \hat{n}_R(X) \right\rangle_{\hat{\rho}}$ for a general state. A significant outcome of this theoretical analysis is that $\left\langle \hat{n}_R(X) \right\rangle_{\hat{\rho}}$ receives non-negligible contributions from left-wedge values of $\psi(x)$. This points toward the possibility for single-particle quantum states to tunnel from the left to the right wedge, across the Rindler horizon. We want to point out that such result does not depend on the particular form of $\psi(x)$. Nevertheless, we have tested the extreme case in which almost all the wave-function is localized beyond the horizon. Specifically, we have considered in detail the case of Gaussian wave-function $\psi(x)$ and verified that in the limit of high locality degree -- i.e. $\sigma/|x_0| \rightarrow 0$ -- the dominant term of $\Delta n_R(X)$ is related to left-wadge values of $\psi(x)$ while the contributions coming from the right tail go to zero exponentially faster. The use of the characteristic function has played a crucial role for deriving the results for single-particle states. Possible generalizations for $\chi[\xi,\xi^*]$ in the case of general Minkowski-Fock states can be obtained through the use of the same identities that have led to Eq. (\ref{characteristic_function_one_particle_final_2}), such as Eq. (\ref{vacuum_state_property}). The development of an explicit form for such characteristic functions will be presented in a future work.

\bibliography{bibliography} 
\bibliographystyle{ieeetr}

\end{document}


\title{Supplemental materials}

\maketitle

\section{(1-1)-dimensional massless scalar fields in Rindler spacetime} \label{massless_scalar_fields_in_Rindler_spacetime}

In this section, we give a brief review of (1-1)-dimensional massless scalar fields in Rindler spacetime. Specifically, we show the explicit form of $\hat{\phi}(t,x)$ -- defined in the main paper as a massless scalar field in Minkowski coordinates -- and its transformed counterparts in Rindler coordinates $\hat{\Phi}_{L,R}(T,X)$. The expression of $\hat{\phi}(t,x)$ ($\hat{\Phi}_{L,R}(T,X)$) will be given in terms of $\hat{a}(k)$ ($ \hat{b}_{L,R}(K)$) defined in the main paper as the annihilation operator for a Minkowski (Rindler) particle with momentum $k$ ($K$). Moreover, we show the explicit expression for the respective conjugate momentum $\hat{\pi}(t,x)$ and $\hat{\Pi}_{L.R}(t,x)$. From the field transformation $\hat{\phi}(t,x) \mapsto \hat{\Phi}_{L,R}(T,X)$ and $\hat{\pi}(t,x) \mapsto \hat{\Pi}_{L,R}(T,X)$ we derive the Bogolyubov transformation $\hat{a}(k) \rightarrow \hat{b}_{L,R}(K)$. Finally we show the canonical commutation relations that define the $\hat{a}(k)$- and $\hat{b}_{L,R}(K)$-algebra.

The massless free scalar field $\hat{\phi}(t,x)$ is solution to the Klein–Gordon equation $\square \hat{\phi}(t,x) = 0$ and writes
\begin{equation} \label{scalar_field_M}
\hat{\phi}(t,x) = \int_{-\infty}^{+\infty} \frac{dk}{\sqrt{2\pi|k|}} \left[ e^{-i|k|ct+ikx} \hat{a}(k) + e^{i|k|ct-ikx}\hat{a}^\dagger (k) \right].
\end{equation}
Analogously, the transformed scalar fields in the Rindler frames $\hat{\Phi}_{L,R}(T,X)$ are still solution to the Klein-Gordon equation and write
\begin{equation} \label{scalar_field_R}
\hat{\Phi}_{L,R}(T,X) =  \int_{-\infty}^{+\infty} \frac{dK}{\sqrt{2\pi|K|}} \left[ e^{-i|K|cT+iKX} \hat{b}_{L,R}(K)  + e^{i|K|cT-iKX} \hat{b}_{L,R}^\dagger(K) \right].
\end{equation}

The conjugate momentum of $\hat{\phi}(t,x)$ in the Minkowski spacetime is defined as $\hat{\pi}(t,x) = c^{-1} \partial_t \hat{\phi}(t,x)$ and explicitly writes
\begin{equation} \label{classical_momentum_field_M}
\hat{\pi}(t,x) =  \int_{-\infty}^{+\infty}dk \sqrt{ \frac{|k|}{2\pi}} i \left[ -e^{-i|k|ct+ikx} \hat{a}(k)  + e^{i|k|ct-ikx} \hat{a}^\dagger(k)\right],
\end{equation}
while in the Rindler spacetime $\hat{\Pi}_{L,R}(T,X) = c^{-1} \partial_T \hat{\Phi}_{L,R}(T,X)$ and hence
\begin{equation} \label{classical_momentum_field_R}
\hat{\Pi}_{L,R}(T,X) =  \int_{-\infty}^{+\infty}dK \sqrt{ \frac{|K|}{2\pi}} i \left[-e^{-i|K|cT+iKX} \hat{b}_{L,R}(K)  + e^{i|K|cT-iKX} \hat{b}^\dagger_{L,R}(K) \right].
\end{equation}

The coefficients $\hat{a}(k)$ can be decomposed as $\hat{a}(k) =  \hat{a}_L(k) + \hat{a}_R(k)$ by defining the fields $\hat{\phi}_{L,R}(x) = \theta(s_{L,R} x)\hat{\phi}(0,x)$, $\hat{\pi}_{L,R}(x) = \theta(s_{L,R} x)\hat{\pi}(0,x)$ and the coefficients $\hat{a}_{L,R}(k)$ such that
\begin{subequations}
\begin{equation}
\hat{\phi}_{L,R}(x) = \int_{-\infty}^{+\infty} \frac{dk}{\sqrt{2\pi|k|}} \left[ e^{ikx} \hat{a}_{L,R}(k) + e^{-ikx}\hat{a}_{L,R}^\dagger(k) \right],
\end{equation}
\begin{equation}
\hat{\pi}_{L,R}(x) =  \int_{-\infty}^{+\infty}dk \sqrt{ \frac{|k|}{2\pi}} i \left[ -e^{ikx} \hat{a}_{L,R}(k)  + e^{-ikx} \hat{a}_{L,R}^\dagger(k)\right].
\end{equation}
\end{subequations}

The Bogoliubov transformations relating $\hat{a}(k)$ and $\hat{b}_{L,R}(K)$ are the following
\begin{equation}\label{Bogolyubov_transformation}
\hat{a}(k) = \int_{-\infty}^{+\infty}dK \left[ \alpha(k,K)\hat{b}_L(K) - \beta^*(k,K) \hat{b}^\dagger_L(K) + \alpha^*(k,K) \hat{b}_R(K) - \beta(k,K) \hat{b}^\dagger_R(K) \right],
\end{equation}
with
\begin{subequations}\label{Bogolyubov_coefficients}
\begin{equation}
\alpha(k,K) = \theta(kK) \sqrt{\frac{K}{k}}F(k,K), \quad \beta(k,K) = \theta(kK) \sqrt{\frac{K}{k}}F(-k,K),
\end{equation}
\begin{equation} \label{F}
F(k,K) = \frac{1}{2 \pi a} \Gamma \left( -\frac{i K}{a} \right) \exp \left( i \frac{K}{a} \ln \frac{|k|}{a} + \text{sign} \left( k \right) \frac{\beta}{4} K \right).
\end{equation}
\end{subequations}
Such transformation can be obtained by requiring that $\hat{\phi}(t,x)$ transforms as a scalar field under any coordinate transformation, while $\hat{\pi}(t,x)$ as a time-derivative of a scalar field. For instance, the coordinate transformation in the right wedge $(t,x) \mapsto (T,X)=(T_R(t,x), X_R(t,x))$ for the fields writes
\begin{subequations}\label{Bogolyubov_transformation_phi_Pi}
\begin{equation}\label{Bogolyubov_transformation_phi}
\hat{\Phi}_R(T,X) = \hat{\phi}(t_R(T,X),x_R(T,X)),
\end{equation}
\begin{align}
\hat{\Pi}_R(T,X) = & \frac{1}{c} \frac{ \partial}{ \partial T} \hat{\Phi}_R(T,X)  \nonumber \\
= & \frac{1}{c} \left( \frac{\partial t_R}{\partial T} \frac{ \partial}{ \partial t} + \frac{\partial x_R}{\partial T}  \frac{ \partial}{ \partial x}  \right) \hat{\phi} (t_R(T, X),x_R(T, X)) \nonumber \\
= & e^{aX} \cosh(a c T) \frac{1}{c}  \frac{ \partial}{ \partial t} \hat{\phi} (t_R(T, X),x_R(T, X)) + e^{aX} \sinh(a c T) \frac{ \partial}{ \partial x} \hat{\phi} (t_R(T, X),x_R(T, X)) \nonumber \\
= & e^{aX} \cosh(a c T) \hat{\pi} (t_R(T, X),x_R(T, X)) + e^{aX} \sinh(a c T) \frac{ \partial}{ \partial x} \hat{\phi} (t_R(T, X),x_R(T, X)).
\end{align}
\end{subequations}
Eqs. (\ref{Bogolyubov_transformation_phi_Pi}) result in the following transformation for $\hat{a}_R(k)$
\begin{align} \label{Bogolyubov_transformation_calculation}
\hat{a}_R(k) = & \int_{-\infty}^{+\infty}dx e^{-ikx} \frac{1}{2} \left[ \sqrt{\frac{|k|}{2\pi}}  \hat{\phi}_R(x) +  \frac{i}{\sqrt{2\pi|k|}} \hat{\pi}_R(x) \right] \nonumber \\
= & \int_{-\infty}^{+\infty}dx e^{-ikx} \frac{1}{2} \left[ \sqrt{\frac{|k|}{2\pi}} \theta(x) \hat{\phi}(0,x) +  \frac{i}{\sqrt{2\pi|k|}} \theta(x) \hat{\pi}(0,x) \right] \nonumber \\
 = & \int_{0}^{+\infty}dx e^{-ikx} \frac{1}{2} \left[ \sqrt{\frac{|k|}{2\pi}}  \hat{\phi}(0,x) +  \frac{i}{\sqrt{2\pi|k|}} \hat{\pi}(0,x) \right] \nonumber \\
 = & \int_{-\infty}^{+\infty}dX e^{a X} \exp (- i k x_R(X)) \frac{1}{2} \left[ \sqrt{\frac{|k|}{2\pi}}  \hat{\phi}(0,x_R(X)) +  \frac{i}{\sqrt{2\pi|k|}} \hat{\pi}(0,x_R(X)) \right] \nonumber \\
 = & \int_{-\infty}^{+\infty}dX \exp \left( -i\frac{k}{a}e^{aX} \right) \frac{1}{2} \left[ \sqrt{\frac{|k|}{2\pi}}  e^{a X}  \hat{\phi}(0,x_R(X)) + \frac{i e^{a X} }{\sqrt{2\pi|k|}} \hat{\pi}(0,x_R(X)) \right] \nonumber \\
= & \int_{-\infty}^{+\infty}dX \exp \left( -i\frac{k}{a}e^{aX} \right) \frac{1}{2} \left[ \sqrt{\frac{|k|}{2\pi}}  e^{aX} \hat{\Phi}_R( 0, X ) +  \frac{i}{\sqrt{2\pi|k|}} \hat{\Pi}_R ( 0, X) \right] \nonumber \\
= & \int_{-\infty}^{+\infty}dX \frac{1}{2} \left[ \sqrt{\frac{|k|}{2\pi}}  \hat{\Phi}_R( 0, X ) ia \frac{d}{dk}  +  \frac{i}{\sqrt{2\pi|k|}} \hat{\Pi}_R ( 0, X) \right] \exp \left( -i\frac{k}{a}e^{aX} \right) \nonumber \\
= & \int_{-\infty}^{+\infty}dX \frac{1}{2} \left\lbrace \sqrt{\frac{|k|}{2\pi}}  \int_{-\infty}^{+\infty} \frac{dK}{\sqrt{2\pi|K|}} \left[ e^{iKX} \hat{b}_R(K)  + e^{-iKX} \hat{b}_R^\dagger(K) \right] ia \frac{d}{dk} \right. \nonumber \\
& \left. - \frac{1}{\sqrt{2\pi|k|}} \int_{-\infty}^{+\infty}dK \sqrt{ \frac{|K|}{2\pi}} \left[-e^{iKX} \hat{b}_R(K)  + e^{-iKX} \hat{b}^\dagger_R(K) \right] \right\rbrace \exp \left( -i\frac{k}{a}e^{aX} \right) \nonumber \\
= & \int_{-\infty}^{+\infty} \frac{dK}{2} \int_{-\infty}^{+\infty}dX \left\lbrace \frac{1}{2\pi} \sqrt{\left|\frac{k}{K}\right|} \left[ e^{iKX} \hat{b}_R(K)  + e^{-iKX} \hat{b}_R^\dagger(K) \right] ia \frac{d}{dk} \right. \nonumber \\
& \left. - \frac{1}{2\pi} \sqrt{\left|\frac{K}{k}\right|} \left[-e^{iKX} \hat{b}_R(K)  + e^{-iKX} \hat{b}^\dagger_R(K) \right] \right\rbrace \exp \left( -i\frac{k}{a}e^{aX} \right) \nonumber \\
 = & \int_{-\infty}^{+\infty}  \frac{dK}{2} \left\lbrace \sqrt{\left|\frac{k}{K}\right|}  \left[ \hat{b}_R(K) i a \frac{d}{dk} F(-k,-K) +  \hat{b}^\dagger_R(K)i a \frac{d}{dk} F(-k,K) \right] \right. \nonumber \\
 & \left. - \sqrt{\left|\frac{K}{k}\right|} \left[  - \hat{b}_R(K) F(-k,-K) + \hat{b}^\dagger_R(K) F(-k,K) \right] \right\rbrace,
\end{align}
where
\begin{equation}
F(k,K) = \int_{-\infty}^{+\infty}\frac{dX}{2\pi}\exp\left(-i K X + i \frac{k}{a} e^{aX} \right)
\end{equation}
is a distribution that can be obtained from the following distributional limit
\begin{equation}\label{F_regularized}
F(k,K) =  \lim_{\epsilon \rightarrow 0^+} \int_{-\infty}^{+\infty}\frac{dX}{2\pi}  \exp\left((-i K + \epsilon a) X + \left(i \frac{k}{a} - \epsilon\right) e^{aX} \right).
\end{equation}
The distribution $F(k,K)$ of Eq. (\ref{F_regularized}) can be proven to be identical to $F(k,K)$ of Eq. (\ref{F}) -- for such proof see for instance Ref. \cite{Mukhanov2007}. The derivative of $F(k,K)$ with respect to $k$ can be obtained by using integration by parts and taking the distributional limit $\epsilon \rightarrow 0^+$:
\begin{align} \label{dFdk}
\frac{d}{dk} F(k,K) = & \lim_{\epsilon \rightarrow 0^+} \int_{-\infty}^{+\infty}dX \frac{i e^{aX}}{2\pi a} \exp\left((-i K + \epsilon a) X + \left(i \frac{k}{a} - \epsilon\right) e^{aX} \right) \nonumber \\
 = & \lim_{\epsilon \rightarrow 0^+} \int_{-\infty}^{+\infty}dX \frac{i}{2\pi a} \exp\left((-i K + \epsilon a) X \right) \frac{1}{a} \left(i \frac{k}{a} - \epsilon\right)^{-1} \frac{d}{dX} \exp \left(\left(i \frac{k}{a} - \epsilon\right) e^{aX} \right) \nonumber \\
= & \lim_{\epsilon \rightarrow 0^+} \left[ \left. \frac{i}{2 \pi a^2} \left(i \frac{k}{a} - \epsilon\right)^{-1} \exp\left((-i K + \epsilon a) X + \left(i \frac{k}{a} - \epsilon\right) e^{aX} \right) \right|_{-\infty}^{+\infty} \right. \nonumber \\
& \left. - \int_{-\infty}^{+\infty}dX\frac{i }{2 \pi a^2} \left(i \frac{k}{a} - \epsilon\right)^{-1} (-iK+\epsilon  a) \exp\left((-i K + \epsilon a) X + \left(i \frac{k}{a} - \epsilon\right) e^{aX} \right) \right] \nonumber \\
= & \lim_{\epsilon \rightarrow 0^+} \left[ \frac{iK-\epsilon a}{a (k + i \epsilon a)} \int_{-\infty}^{+\infty} \frac{dX}{2 \pi} \exp\left((-i K + \epsilon a) X + \left(i \frac{k}{a} - \epsilon\right) e^{aX} \right) \right] \nonumber \\
= & \frac{iK}{ak} F(k,K).
\end{align}
In this way the calculation of Eq. (\ref{Bogolyubov_transformation_calculation}) leads to
\begin{align}\label{Bogolyubov_transformation_calculation_2}
\hat{a}_R(k)  = & \int_{-\infty}^{+\infty}  \frac{dK}{2} \left\lbrace \sqrt{\left|\frac{k}{K}\right|}  \left[ \hat{b}_R(K)  \frac{K}{k} F(-k,-K) -  \hat{b}^\dagger_R(K) \frac{K}{k} F(-k,K) \right] \right. \nonumber \\
 & \left. - \sqrt{\left|\frac{K}{k}\right|} \left[  - \hat{b}_R(K) F(-k,-K) + \hat{b}^\dagger_R(K) F(-k,K) \right] \right\rbrace \nonumber \\
   = & \int_{-\infty}^{+\infty}  \frac{dK}{2} \left\lbrace \sqrt{\left|\frac{k}{K}\right|}  \left[ \hat{b}_R(K)  \text{sign}(kK) \left| \frac{K}{k} \right| F(-k,-K) -  \hat{b}^\dagger_R(K) \text{sign}(kK) \left| \frac{K}{k} \right| F(-k,K) \right] \right. \nonumber \\
 & \left. - \sqrt{\left|\frac{K}{k}\right|} \left[  - \hat{b}_R(K) F(-k,-K) + \hat{b}^\dagger_R(K) F(-k,K) \right] \right\rbrace \nonumber \\
   = & \int_{-\infty}^{+\infty}  \frac{dK}{2} \sqrt{\left|\frac{K}{k}\right|}   \left[ \hat{b}_R(K)  \text{sign}(kK) F(-k,-K) -  \hat{b}^\dagger_R(K) \text{sign}(kK) F(-k,K)  \right. \nonumber \\
 & \left. + \hat{b}_R(K) F(-k,-K) - \hat{b}^\dagger_R(K) F(-k,K) \right]  \nonumber \\
  = & \int_{-\infty}^{+\infty}dK \sqrt{\left|\frac{K}{k}\right|} \left[  \frac{ 1 + \text{sign}(kK) }{2} F(-k,-K) \hat{b}_R(K) - \frac{ 1 + \text{sign}(kK) }{2} F(-k,K) \hat{b}_R^\dagger(K) \right] \nonumber \\
    = & \int_{-\infty}^{+\infty}dK \sqrt{\left|\frac{K}{k}\right|} \left[  \theta(kK) F(-k,-K) \hat{b}_R(K) - \theta(kK) F(-k,K) \hat{b}_R^\dagger(K) \right] \nonumber \\
= & \int_{-\infty}^{+\infty}dK \left[ \alpha^*(k,K) \hat{b}_R(K) - \beta(k,K) \hat{b}^\dagger_R(K) \right].
\end{align}
Moreover the transformation $a\mapsto-a$ is equivalent to $F(k,K) \mapsto F^*(k,K)$. In this way we can easily prove the following identity from Eq. (\ref{Bogolyubov_transformation_calculation_2})
\begin{equation}\label{Bogolyubov_transformation_calculation_3}
\hat{a}_L(k) = \int_{-\infty}^{+\infty}dK \left[ \alpha(k,K) \hat{b}_L(K) - \beta^*(k,K) \hat{b}^\dagger_L(K) \right].
\end{equation}
Eqs. (\ref{Bogolyubov_transformation_calculation_2}, \ref{Bogolyubov_transformation_calculation_3}) result in Eq. (\ref{Bogolyubov_transformation}).

The inverse equations for Eq. (\ref{Bogolyubov_transformation}) read
\begin{equation}\label{Bogolyubov_transformation_inverse}
b_L(K) = \int_{-\infty}^{+\infty}dk \left[ \alpha^*(k, K)\hat{a}(k) + \beta^*(k,K)\hat{a}^\dagger(k) \right], \quad \hat{b}_R(K) = \int_{-\infty}^{+\infty}dk \left[ \alpha(k, K)\hat{a}(k) + \beta(k,K)\hat{a}^\dagger(k) \right].
\end{equation}
Eq. (\ref{Bogolyubov_transformation_inverse}) will be proven in Sec. \ref{Properties_of_the_Bogolyubov_coefficients}.

Any quantum field in Minkowski space-time must be provided with the following canonical commutation relation 
\begin{equation}\label{canonical_commutation_Minkowski}
\left[\hat{a}(k),\hat{a}(k')\right]=0, \quad 
\left[\hat{a}(k),\hat{a}^\dagger(k')\right]=\delta(k-k').
\end{equation}
The analogue procedure in the Rindler spacetime consists into imposing the following commutation relations
\begin{equation}\label{canonical_commutation_Rindler}
\left[\hat{b}_L(K),\hat{b}_L(K')\right]=0, \quad \left[\hat{b}_L(K),\hat{b}_L^\dagger(K')\right]=\delta(K-K'), \quad \left[\hat{b}_R(K),\hat{b}_R(K')\right]=0, \quad \left[\hat{b}_R(K),\hat{b}_R^\dagger(K')\right]=\delta(K-K').
\end{equation}
It is possible to prove that the commutation relations (\ref{canonical_commutation_Minkowski}) are compatible with (\ref{canonical_commutation_Rindler}) and give the following additional commutation relations
\begin{equation}\label{canonical_commutation_Rindler_LR}
\left[\hat{b}_L(K),\hat{b}_R(K')\right]=0, \quad \left[\hat{b}_L(K),\hat{b}_R^\dagger(K')\right]=0.
\end{equation}
See Sec. \ref{Properties_of_the_Bogolyubov_coefficients} for such proof.

\section{Properties of the Bogolyubov coefficients}\label{Properties_of_the_Bogolyubov_coefficients}

In Sec. \ref{massless_scalar_fields_in_Rindler_spacetime} we have derived the Bogoliubov transformations relating $\hat{a}(k)$ and $\hat{b}_{L,R}(K)$. The explicit expression for the Bogoliubov coefficients is shown by Eqs. (\ref{Bogolyubov_coefficients}). From such equations we can extract some identities which appear to be useful for some proofs.

For instance, Eq. (\ref{Bogolyubov_transformation_inverse}) can be proven through the following identities 
\begin{subequations} \label{identity_k}
\begin{equation} \label{identity_k_3}
\int_{-\infty}^{+\infty}dk [ \alpha(k,K) \alpha(k,K') - \beta(k,K)\beta(k,K') ] = 0,
\end{equation}
\begin{equation} \label{identity_k_4}
\int_{-\infty}^{+\infty}dk [ \alpha(k,K) \beta^*(k,K') - \beta(k,K)\alpha^*(k,K') ] = 0,
\end{equation}
\begin{equation} \label{identity_k_1}
\int_{-\infty}^{+\infty}dk [ \alpha(k,K) \beta(k,K') - \beta(k,K)\alpha(k,K') ] = 0,
\end{equation}
\begin{equation} \label{identity_k_2}
\int_{-\infty}^{+\infty}dk [ \alpha(k,K) \alpha^*(k,K') - \beta(k,K)\beta^*(k,K') ] = \delta(K-K'),
\end{equation}
\end{subequations}
which are valid for any $K, K' \in \mathbb{R}$. Indeed, Eqs. (\ref{identity_k}) and their conjugates result in
\begin{align}
& \int_{-\infty}^{+\infty}dk \left[ \alpha^*(k, K)\hat{a}(k) + \beta^*(k,K)\hat{a}^\dagger(k) \right] \nonumber \\
= & \int_{-\infty}^{+\infty}dk \left\lbrace \alpha^*(k, K) \int_{-\infty}^{+\infty}dK' \left[ \alpha(k,K')\hat{b}_L(K') - \beta^*(k,K') \hat{b}^\dagger_L(K') + \alpha^*(k,K') \hat{b}_R(K') - \beta(k,K') \hat{b}^\dagger_R(K') \right] \right. \nonumber \\
& \left. + \beta^*(k,K) \int_{-\infty}^{+\infty}dK' \left[ \alpha^*(k,K')\hat{b}^\dagger_L(K') - \beta(k,K') \hat{b}_L(K') + \alpha(k,K') \hat{b}^\dagger_R(K') - \beta^*(k,K') \hat{b}_R(K') \right] \right\rbrace \nonumber \\
= & \int_{-\infty}^{+\infty}dK' \int_{-\infty}^{+\infty}dk \left[  \alpha^*(k, K) \alpha(k,K')\hat{b}_L(K') -  \alpha^*(k, K) \beta^*(k,K') \hat{b}^\dagger_L(K') +  \alpha^*(k, K) \alpha^*(k,K') \hat{b}_R(K') \right. \nonumber \\
& -  \alpha^*(k, K) \beta(k,K')  \hat{b}^\dagger_R(K') +  \beta^*(k,K) \alpha^*(k,K')\hat{b}^\dagger_L(K') -  \beta^*(k,K) \beta(k,K') \hat{b}_L(K') +  \beta^*(k,K) \alpha(k,K') \hat{b}^\dagger_R(K')  \nonumber \\
& \left. - \beta^*(k,K) \beta^*(k,K') \hat{b}_R(K') \right]  \nonumber \\
= & \int_{-\infty}^{+\infty}dK' \int_{-\infty}^{+\infty}dk \left\lbrace [ \alpha^*(k, K) \alpha(k,K') -  \beta^*(k,K) \beta(k,K') ] \hat{b}_L(K') + [ \beta^*(k,K) \alpha^*(k,K') -  \alpha^*(k, K) \beta^*(k,K')  ] \right. \nonumber \\
& \left. \times \hat{b}^\dagger_L(K') + [\alpha^*(k, K) \alpha^*(k,K')-  \beta^*(k,K) \beta^*(k,K')] \hat{b}_R(K') + [ \beta^*(k,K) \alpha(k,K') -  \alpha^*(k, K) \beta(k,K')] \hat{b}^\dagger_R(K')  \right\rbrace   \nonumber \\
= & \int_{-\infty}^{+\infty}dK' \delta(K-K') \hat{b}_L(K') \nonumber \\
= & \hat{b}_L(K)
\end{align}
and
\begin{align}
& \int_{-\infty}^{+\infty}dk \left[ \alpha(k, K)\hat{a}(k) + \beta(k,K)\hat{a}^\dagger(k) \right] \nonumber \\
= & \int_{-\infty}^{+\infty}dk \left\lbrace \alpha(k, K) \int_{-\infty}^{+\infty}dK' \left[ \alpha(k,K')\hat{b}_L(K') - \beta^*(k,K') \hat{b}^\dagger_L(K') + \alpha^*(k,K') \hat{b}_R(K') - \beta(k,K') \hat{b}^\dagger_R(K') \right] \right. \nonumber \\
& \left. + \beta(k,K) \int_{-\infty}^{+\infty}dK' \left[ \alpha^*(k,K')\hat{b}^\dagger_L(K') - \beta(k,K') \hat{b}_L(K') + \alpha(k,K') \hat{b}^\dagger_R(K') - \beta^*(k,K') \hat{b}_R(K') \right] \right\rbrace \nonumber \\
= & \int_{-\infty}^{+\infty}dK' \int_{-\infty}^{+\infty}dk \left[  \alpha(k, K) \alpha(k,K')\hat{b}_L(K') -  \alpha(k, K) \beta^*(k,K') \hat{b}^\dagger_L(K') +  \alpha(k, K) \alpha^*(k,K') \hat{b}_R(K') \right. \nonumber \\
& -  \alpha(k, K) \beta(k,K')  \hat{b}^\dagger_R(K') +  \beta(k,K) \alpha^*(k,K')\hat{b}^\dagger_L(K') -  \beta(k,K) \beta(k,K') \hat{b}_L(K') +  \beta(k,K) \alpha(k,K') \hat{b}^\dagger_R(K')  \nonumber \\
& \left. - \beta(k,K) \beta^*(k,K') \hat{b}_R(K') \right]  \nonumber \\
= & \int_{-\infty}^{+\infty}dK' \int_{-\infty}^{+\infty}dk \left\lbrace [ \alpha(k, K) \alpha(k,K') -  \beta(k,K) \beta(k,K') ] \hat{b}_L(K') + [ \beta(k,K) \alpha^*(k,K') -  \alpha(k, K) \beta^*(k,K')  ] \right. \nonumber \\
& \left. \times \hat{b}^\dagger_L(K') + [\alpha(k, K) \alpha^*(k,K')-  \beta(k,K) \beta^*(k,K')] \hat{b}_R(K') + [ \beta(k,K) \alpha(k,K') -  \alpha(k, K) \beta(k,K')] \hat{b}^\dagger_R(K')  \right\rbrace   \nonumber \\
= & \int_{-\infty}^{+\infty}dK' \delta(K-K') \hat{b}_R(K') \nonumber \\
= & \hat{b}_R(K).
\end{align}
Moreover, Eq. (\ref{Bogolyubov_transformation_inverse}) and Eqs. (\ref{identity_k}) can be used in order to prove  Eqs. (\ref{canonical_commutation_Rindler}, \ref{canonical_commutation_Rindler_LR}) from Eq. (\ref{canonical_commutation_Minkowski}).

Eqs. (\ref{identity_k}) come from the following chains of identities valid for any $p, q \in \{ 1, -1 \}$
\begin{align} \label{identity_k_5}
&\int_{-\infty}^{+\infty}dk \theta(kK) \theta(kK') \frac{\sqrt{KK'}}{|k|} \left[ F(k,K)F(-pk,-qK') - F(-k,K)F(pk,-qK')  \right] \nonumber \\
= & \int_{-\infty}^{+\infty}dk \theta(kK)\theta(KK') \frac{\sqrt{KK'}}{|k|} \frac{1}{(2 \pi a)^2} \Gamma \left( -\frac{i K}{a} \right) \Gamma \left( q \frac{i K'}{a} \right)  \left[ \exp \left( i \frac{K-qK'}{a} \ln \frac{|k|}{a} + \text{sign} (k) \frac{\beta}{4} ( K +  p q K') \right) \right. \nonumber \\
& \left. - \exp \left( i \frac{K-qK'}{a} \ln \frac{|k|}{a} - \text{sign} (k) \frac{\beta}{4} (K + p q  K') \right)  \right] \nonumber \\
= & \theta(KK') \sqrt{KK'} \Gamma \left( -\frac{i K}{a} \right) \Gamma \left( q \frac{i K'}{a} \right) \int_{-\infty}^{+\infty}dk \frac{\theta(kK)}{|k|} \frac{1}{(2 \pi a)^2}  \left[ \exp \left( i \frac{K-qK'}{a} \ln \frac{|k|}{a} + \text{sign} (K) \frac{\beta}{4} ( K +  p q K') \right) \right. \nonumber \\
& \left. - \exp \left( i \frac{K-qK'}{a} \ln \frac{|k|}{a} - \text{sign} (K) \frac{\beta}{4} (K + p q  K') \right)  \right] \nonumber \\
= & \frac{\theta(KK') \sqrt{KK'}}{2 \pi a} \Gamma \left( -\frac{i K}{a} \right) \Gamma \left( q \frac{i K'}{a} \right) 2 \sinh \left( \text{sign} (K) \frac{\beta}{4} ( K +  p q K') \right) \int_{-\infty}^{+\infty}dk \frac{\theta(kK)}{2 \pi a|k|}  \exp \left( i \frac{K-qK'}{a} \ln \frac{|k|}{a} \right)  \nonumber \\
= & \frac{\theta(KK') \sqrt{KK'}}{ \pi a} \Gamma \left( -\frac{i K}{a} \right) \Gamma \left( q \frac{i K'}{a} \right) \sinh \left( \text{sign} (K) \frac{\beta}{4} ( K +  p q K') \right) \int_{0}^{+\infty} \frac{d|k|}{2 \pi a|k|}  \exp \left( i \frac{K-qK'}{a} \ln \frac{|k|}{a} \right)  \nonumber \\
= & \frac{\theta(KK') \sqrt{KK'}}{\pi a} \Gamma \left( -\frac{i K}{a} \right) \Gamma \left( q \frac{i K'}{a} \right) \sinh \left( \text{sign} (K) \frac{\beta}{4} ( K +  p q K') \right) \int_{-\infty}^{+\infty}\frac{d\Xi}{2 \pi}  \exp \left( i (K-qK') \Xi \right)  \nonumber \\
= & \frac{\theta(KK') \sqrt{KK'}}{\pi a} \Gamma \left( -\frac{i K}{a} \right) \Gamma \left( q \frac{i K'}{a} \right) \sinh \left( \text{sign} (K) \frac{\beta}{4} ( K +  p q K') \right) \delta(K-qK')  \nonumber \\
= & \frac{\theta(q) |K|}{\pi a} \left| \Gamma \left( \frac{i K}{a} \right) \right|^2 \sinh \left( \text{sign}(K) \frac{\beta}{4} (1+p) K \right) \delta(K-qK')   \nonumber \\
= & \theta(q) \left[ \sinh \left( \frac{\beta}{2} |K| \right) \right]^{-1} \sinh \left( \frac{\beta}{4} (1+p) |K| \right) \delta(K-qK'),
\end{align}
where we have replaced the integration variable $|k|$ with
\begin{equation}
\Xi = \frac{1}{a} \ln \left(\frac{|k|}{a}\right)
\end{equation}
and used the following equation for the gamma function
\begin{equation}
|\Gamma(i z)|^2 = \frac{\pi}{|z| \sinh(\pi |z|)},
\end{equation}
which is valid for any real $z$. For $p=-1$ Eq. (\ref{identity_k_5}) gives
\begin{equation}
\int_{-\infty}^{+\infty}dk \theta(kK) \theta(kK')  \frac{\sqrt{KK'}}{|k|} \left[ F(k,K)F(k,-qK')  - F(-k,K)F(-k,-qK')  \right] = 0,
\end{equation}
resulting in Eqs. (\ref{identity_k_3}, \ref{identity_k_4}), while for $p=1$ Eq. (\ref{identity_k_5}) gives
\begin{equation} \label{identity_k_6}
\int_{-\infty}^{+\infty}dk \theta(kK) \theta(kK')  \frac{\sqrt{KK'}}{|k|} \left[ F(k,K)F(-k,-qK') - F(-k,K)F(k,-qK')  \right] = \theta(q) \delta(K-qK'),
\end{equation}
For $q=-1$ Eq. (\ref{identity_k_6}) gives
\begin{equation}
\int_{-\infty}^{+\infty}dk \theta(kK) \theta(kK')  \frac{\sqrt{KK'}}{|k|} \left[ F(k,K)F(-k,K')  - F(-k,K)F(k,K')  \right] = 0,
\end{equation}
resulting in Eq. (\ref{identity_k_1}), while for $q=1$ Eq. (\ref{identity_k_6}) gives
\begin{equation}
\int_{-\infty}^{+\infty}dk \theta(kK) \theta(kK')  \frac{\sqrt{KK'}}{|k|} \left[ F(k,K)F(-k,-K') - F(-k,K)F(k,-K')  \right] = \delta(K-K'),
\end{equation}
resulting in Eq. (\ref{identity_k_2}).

Other important identities relating $\alpha(k,K)$ and $\beta(k,K)$ are the following:
\begin{equation} \label{alpha_to_beta}
\alpha(k,K) = \exp \left( \frac{\beta}{2} |K| \right) \beta(k,K) ,
\end{equation}
\begin{equation} \label{identity_K}
 \int_{-\infty}^{+\infty}dK 2 \sinh \left( \frac{\beta}{2} |K| \right) \left[ \alpha(k,K)\beta^*(k',K) + \beta^*(k,K)\alpha(k',K)  \right] = \delta(k-k').
\end{equation}
Eq. (\ref{alpha_to_beta}) comes from the following identity
\begin{equation} \label{alpha_beta_to_alpha_beta_proof}
F(k,K) = \exp \left( \text{sign}(k) \frac{\beta}{2} K \right) F(-k,K),
\end{equation}
which can be directly extracted from Eq. (\ref{F}). On the other hand, Eq. (\ref{identity_K}) can be proven by the following chain of identities:
\begin{align}
& \int_{-\infty}^{+\infty}dK \theta(kK) \theta(k'K) 2 \sinh \left( \frac{\beta}{2} |K| \right) \frac{|K|}{\sqrt{kk'}}  \left[ F(k,K)F(k',-K) + F(k,-K)F(k',K)  \right] \nonumber \\
= &  \frac{\theta(kk')}{\sqrt{kk'}} \left[ \int_{-\infty}^{+\infty}dK \theta(kK) 2 \sinh \left( \frac{\beta}{2} |K| \right) |K|  F(k,K)F(k',-K) \right. \nonumber \\
& \left. + \int_{-\infty}^{+\infty}dK \theta(kK) 2 \sinh \left( \frac{\beta}{2} |K| \right) |K| F(k,-K)F(k',K)  \right] \nonumber \\
= &  \frac{\theta(kk')}{\sqrt{kk'}} \left[ \int_{-\infty}^{+\infty}dK \theta(kK) 2 \sinh \left( \frac{\beta}{2} |K| \right) |K|  F(k,K)F(k',-K) \right. \nonumber \\
& \left. + \int_{-\infty}^{+\infty}dK \theta(-kK) 2 \sinh \left( \frac{\beta}{2} |K| \right) |K| F(k,K)F(k',-K)  \right] \nonumber \\
= &  \frac{\theta(kk')}{\sqrt{kk'}} \int_{-\infty}^{+\infty}dK \left[ \theta(kK)+\theta(-kK) \right] 2 \sinh \left( \frac{\beta}{2} |K| \right) |K|  F(k,K)F(k',-K) \nonumber \\
= &  \frac{\theta(kk')}{\sqrt{kk'}} \int_{-\infty}^{+\infty}dK 2 \sinh \left( \frac{\beta}{2} |K| \right) |K|  F(k,K)F(k',-K) \nonumber \\
= & \frac{\theta(kk')}{\sqrt{kk'}} \int_{-\infty}^{+\infty}dK 2 \sinh \left( \frac{\beta}{2} |K| \right) \frac{|K|}{(2 \pi a)^2} \left| \Gamma \left( \frac{i K}{a} \right) \right|^2 \exp \left( i \frac{K}{a} \ln \left| \frac{k}{k'} \right| \right) \nonumber \\
= & \frac{\theta(kk')}{\sqrt{kk'}} \int_{-\infty}^{+\infty} \frac{dK}{2 \pi a}  \exp \left( i \frac{K}{a} \ln \left| \frac{k}{k'} \right| \right)\nonumber \\
= & \frac{\theta(kk')}{\sqrt{kk'}} \delta \left( \ln \left| \frac{k}{k'} \right| \right) \nonumber \\
= & \frac{\theta(kk')}{|k|} \delta \left( \ln \left| \frac{k}{k'} \right| \right) \nonumber \\
= & \theta(kk') \delta \left( |k| - |k'| \right)\nonumber \\
= & \delta \left( k - k' \right).
\end{align}

\section{Minkowski vacuum states}

In this section, we give a brief review of the Minkowski vacuum states seen by an accelerated observer. The aim is to provide a simple proof for the Rindler effect through some identities shown in the previous sections. Moreover, we provide a proof for Eq. (5) of the main paper and for some other identities which will be used in the next sections.

The Minkowski vacuum states is defined by
\begin{equation} \label{vacuum_state_H_M_in_tilde_H'_M_epsilon}
\hat{a}(k) |0_M\rangle = 0, \quad \forall k \in \mathbb{R}.
\end{equation}
Eq. (\ref{vacuum_state_H_M_in_tilde_H'_M_epsilon}) supplemented with Eqs. (\ref{Bogolyubov_transformation}, \ref{alpha_to_beta}) gives
\begin{equation}
\int_{-\infty}^{+\infty}dK \left\lbrace\beta(k,K) \left[ \exp \left( \frac{\beta}{2} |K| \right)  \hat{b}_L(K) -  \hat{b}^\dagger_R(K) \right] + \beta^*(k,K) \left[\exp \left( \frac{\beta}{2} |K| \right) \hat{b}_R(K) -  \hat{b}^\dagger_L(K) \right] \right\rbrace|0_M\rangle = 0,
\end{equation}
which holds for any real $k$ and resulting, therefore, in Eq. (5).

A non-normalizable state which satisfy Eq. (5) exists and its explicit expression is the following
\begin{equation}  \label{vacuum_state_H_M_in_H'_M}
|0_M\rangle \propto \hat{S} | 0_L, 0_R \rangle,
\end{equation}
with
\begin{equation} \label{D}
\hat{S} = \exp \left( \int_{-\infty}^{+\infty} dK \exp \left( -\frac{\beta}{2} |K| \right) \hat{b}^\dagger_L(K) \hat{b}^\dagger_R(K) \right).
\end{equation}
Eq. (\ref{vacuum_state_H_M_in_H'_M}) is identical to Eq. (1) of the main paper and, therefore, it results in the Unruh effect. The fact that the $|0_M\rangle$ in Eq. (\ref{vacuum_state_H_M_in_H'_M}) satisfies Eqs. (5) can be proven in the following way
\begin{align}
\left[ \hat{b}_{L,R}(K) , \hat{S} \right] = & \left[ \hat{b}_{L,R}(K) , \hat{1} +\int_{-\infty}^{+\infty} dK_1 \exp \left( -\frac{\beta}{2} |K_1| \right) \hat{b}^\dagger_L(K_1) \hat{b}^\dagger_R(K_1) +  \frac{1}{2}\int_{-\infty}^{+\infty} dK_1 \exp \left( -\frac{\beta}{2} |K_1| \right) \hat{b}^\dagger_L(K_1) \hat{b}^\dagger_R(K_1) \right. \nonumber \\
& \left. \times \int_{-\infty}^{+\infty} dK_2 \exp \left( -\frac{\beta}{2} |K_2| \right) \hat{b}^\dagger_L(K_2) \hat{b}^\dagger_R(K_2) + \sum_{n=3}^{\infty} \frac{1}{n!} \prod_{i=1}^{n}   \int_{-\infty}^{+\infty} dK_i \exp \left( -\frac{\beta}{2} |K_i| \right) \hat{b}^\dagger_L(K_i) \hat{b}^\dagger_R(K_i) \right] \nonumber \\
 = & \left[ \hat{b}_{L,R}(K) , \hat{1} \right] + \int_{-\infty}^{+\infty} dK_1 \exp \left( -\frac{\beta}{2} |K_1| \right) \left[ \hat{b}_{L,R}(K) , \hat{b}^\dagger_L(K_1) \hat{b}^\dagger_R(K_1) \right] +  \frac{1}{2} \int_{-\infty}^{+\infty} dK_1 \exp \left( -\frac{\beta}{2} |K_1| \right)  \nonumber \\
&  \times \left[ \hat{b}_{L,R}(K) ,\hat{b}^\dagger_L(K_1) \hat{b}^\dagger_R(K_1)\right] \int_{-\infty}^{+\infty} dK_2 \exp \left( -\frac{\beta}{2} |K_2| \right) \hat{b}^\dagger_L(K_2) \hat{b}^\dagger_R(K_2)  +  \frac{1}{2} \int_{-\infty}^{+\infty} dK_1 \exp \left( -\frac{\beta}{2} |K_1| \right) \nonumber \\
&  \times  \hat{b}^\dagger_L(K_1) \hat{b}^\dagger_R(K_1) \int_{-\infty}^{+\infty} dK_2 \exp \left( -\frac{\beta}{2} |K_2| \right) \left[ \hat{b}_{L,R}(K) ,\hat{b}^\dagger_L(K_2) \hat{b}^\dagger_R(K_2) \right] + \sum_{n=3}^{\infty} \frac{1}{n!} \left\lbrace \int_{-\infty}^{+\infty} dK_1 \right. \nonumber \\
 &  \times  \exp \left( -\frac{\beta}{2} |K_1| \right)\left[ \hat{b}_{L,R}(K) , \hat{b}^\dagger_L(K_1) \hat{b}^\dagger_R(K_1) \right]  \prod_{i=2}^{n}  \int_{-\infty}^{+\infty} dK_i \exp \left( -\frac{\beta}{2} |K_i| \right) \hat{b}^\dagger_L(K_i) \hat{b}^\dagger_R(K_i)  \nonumber \\
 &   + \sum_{j=2}^{n-1}  \prod_{i=1}^{j-1}  \int_{-\infty}^{+\infty} dK_i \exp \left( -\frac{\beta}{2} |K_i| \right) \hat{b}^\dagger_L(K_i) \hat{b}^\dagger_R(K_i) \int_{-\infty}^{+\infty} dK_j \exp \left( -\frac{\beta}{2} |K_j| \right) \left[ \hat{b}_{L,R}(K) , \hat{b}^\dagger_L(K_j) \right.   \nonumber \\
 &  \left. \times \hat{b}^\dagger_R(K_j) \right] \prod_{i=j+1}^{n}  \int_{-\infty}^{+\infty} dK_i \exp \left( -\frac{\beta}{2} |K_i| \right) \hat{b}^\dagger_L(K_i) \hat{b}^\dagger_R(K_i)  +  \prod_{i=1}^{n-1}  \int_{-\infty}^{+\infty} dK_i \exp \left( -\frac{\beta}{2} |K_i| \right) \nonumber \\
 & \left. \times \hat{b}^\dagger_L(K_i) \hat{b}^\dagger_R(K_i)  \int_{-\infty}^{+\infty} dK_n \exp \left( -\frac{\beta}{2} |K_n| \right)  \left[ \hat{b}_{L,R}(K) , \hat{b}^\dagger_L(K_n) \hat{b}^\dagger_R(K_n) \right] \right\rbrace\nonumber \\
  = &  \int_{-\infty}^{+\infty} dK_1 \exp \left( -\frac{\beta}{2} |K_1| \right) \hat{b}^\dagger_{R,L}(K_1) \delta(K-K_1) + \frac{1}{2} \int_{-\infty}^{+\infty} dK_1 \exp \left( -\frac{\beta}{2} |K_1| \right)  \hat{b}^\dagger_{R,L}(K_1) \delta(K-K_1)  \nonumber \\
&  \times\int_{-\infty}^{+\infty} dK_2 \exp \left( -\frac{\beta}{2} |K_2| \right) \hat{b}^\dagger_L(K_2) \hat{b}^\dagger_R(K_2)  +  \frac{1}{2} \int_{-\infty}^{+\infty} dK_1 \exp \left( -\frac{\beta}{2} |K_1| \right) \hat{b}^\dagger_L(K_1) \hat{b}^\dagger_R(K_1) \int_{-\infty}^{+\infty} dK_2  \nonumber \\
&  \times \exp \left( -\frac{\beta}{2} |K_2| \right) \hat{b}^\dagger_{R,L}(K_2) \delta(K-K_2) + \sum_{n=3}^{\infty} \frac{1}{n!} \left[ \int_{-\infty}^{+\infty} dK_1 \exp \left( -\frac{\beta}{2} |K_1| \right) \hat{b}^\dagger_{R,L}(K_1) \delta(K-K_1) \right. \nonumber \\
 &  \times \prod_{i=2}^{n}  \int_{-\infty}^{+\infty} dK_i \exp \left( -\frac{\beta}{2} |K_i| \right) \hat{b}^\dagger_L(K_i) \hat{b}^\dagger_R(K_i)  + \sum_{j=2}^{n-1}  \prod_{i=1}^{j-1}  \int_{-\infty}^{+\infty} dK_i \exp \left( -\frac{\beta}{2} |K_i| \right) \hat{b}^\dagger_L(K_i) \hat{b}^\dagger_R(K_i)   \nonumber \\
 &  \times  \int_{-\infty}^{+\infty} dK_j \exp \left( -\frac{\beta}{2} |K_j| \right) \hat{b}^\dagger_{R,L}(K_j) \delta(K-K_j)  \prod_{i=j+1}^{n}  \int_{-\infty}^{+\infty} dK_i \exp \left( -\frac{\beta}{2} |K_i| \right) \hat{b}^\dagger_L(K_i) \hat{b}^\dagger_R(K_i)  \nonumber \\
 & \left.   +  \prod_{i=1}^{n-1}  \int_{-\infty}^{+\infty} dK_i \exp \left( -\frac{\beta}{2} |K_i| \right) \hat{b}^\dagger_L(K_i) \hat{b}^\dagger_R(K_i)  \int_{-\infty}^{+\infty} dK_n \exp \left( -\frac{\beta}{2} |K_n| \right)  \hat{b}^\dagger_{R,L}(K_n) \delta(K-K_n)  \right]\nonumber \\
  = &  \exp \left( -\frac{\beta}{2} |K| \right) \hat{b}^\dagger_{R,L}(K) +  \frac{1}{2} \exp \left( -\frac{\beta}{2} |K| \right)  \hat{b}^\dagger_{R,L}(K)  \int_{-\infty}^{+\infty} dK_2 \exp \left( -\frac{\beta}{2} |K_2| \right) \hat{b}^\dagger_L(K_2) \hat{b}^\dagger_R(K_2)  \nonumber \\
&   +  \frac{1}{2} \int_{-\infty}^{+\infty} dK_1 \exp \left( -\frac{\beta}{2} |K_1| \right) \hat{b}^\dagger_L(K_1) \hat{b}^\dagger_R(K_1)  \exp \left( -\frac{\beta}{2} |K| \right) \hat{b}^\dagger_{R,L}(K) + \sum_{n=3}^{\infty} \frac{1}{n!} \left[ \exp \left( -\frac{\beta}{2} |K| \right) \right.\nonumber \\
&  \times  \hat{b}^\dagger_{R,L}(K) \prod_{i=2}^{n}  \int_{-\infty}^{+\infty} dK_i \exp \left( -\frac{\beta}{2} |K_i| \right) \hat{b}^\dagger_L(K_i) \hat{b}^\dagger_R(K_i)  + \sum_{j=2}^{n-1}  \prod_{i=1}^{j-1}  \int_{-\infty}^{+\infty} dK_i \exp \left( -\frac{\beta}{2} |K_i| \right)   \nonumber \\
 & \times   \hat{b}^\dagger_L(K_i) \hat{b}^\dagger_R(K_i) \exp \left( -\frac{\beta}{2} |K| \right) \hat{b}^\dagger_{R,L}(K)   \prod_{i=j+1}^{n}  \int_{-\infty}^{+\infty} dK_i \exp \left( -\frac{\beta}{2} |K_i| \right)   \hat{b}^\dagger_L(K_i) \hat{b}^\dagger_R(K_i) \nonumber \\
 & \left.   +  \prod_{i=1}^{n-1}  \int_{-\infty}^{+\infty} dK_i \exp \left( -\frac{\beta}{2} |K_i| \right) \hat{b}^\dagger_L(K_i) \hat{b}^\dagger_R(K_i) \exp \left( -\frac{\beta}{2} |K| \right)  \hat{b}^\dagger_{R,L}(K) \right] \nonumber \\
   = &  \exp \left( -\frac{\beta}{2} |K| \right) \hat{b}^\dagger_{R,L}(K) \left\lbrace 1 +  \frac{1}{2}  \int_{-\infty}^{+\infty} dK_2 \exp \left( -\frac{\beta}{2} |K_2| \right) \hat{b}^\dagger_L(K_2) \hat{b}^\dagger_R(K_2)  +  \frac{1}{2} \int_{-\infty}^{+\infty} dK_1 \right. \nonumber \\
&  \times \exp \left( -\frac{\beta}{2} |K_1| \right) \hat{b}^\dagger_L(K_1) \hat{b}^\dagger_R(K_1)  + \sum_{n=3}^{\infty} \frac{1}{n!} \left[  \prod_{i=2}^{n}  \int_{-\infty}^{+\infty} dK_i \exp \left( -\frac{\beta}{2} |K_i| \right) \hat{b}^\dagger_L(K_i) \hat{b}^\dagger_R(K_i) \right.\nonumber \\
&   + \sum_{j=2}^{n-1}  \prod_{i=1}^{j-1}  \int_{-\infty}^{+\infty} dK_i \exp \left( -\frac{\beta}{2} |K_i| \right)   \hat{b}^\dagger_L(K_i) \hat{b}^\dagger_R(K_i)  \prod_{i=j+1}^{n}  \int_{-\infty}^{+\infty} dK_i \exp \left( -\frac{\beta}{2} |K_i| \right)   \hat{b}^\dagger_L(K_i) \hat{b}^\dagger_R(K_i) \nonumber \\
 & \left. \left.  +  \prod_{i=1}^{n-1}  \int_{-\infty}^{+\infty} dK_i \exp \left( -\frac{\beta}{2} |K_i| \right) \hat{b}^\dagger_L(K_i) \hat{b}^\dagger_R(K_i) \right] \right\rbrace \nonumber \\
    = &  \exp \left( -\frac{\beta}{2} |K| \right) \hat{b}^\dagger_{R,L}(K) \left\lbrace 1 +  \int_{-\infty}^{+\infty} dK' \exp \left( -\frac{\beta}{2} |K'| \right) \hat{b}^\dagger_L(K') \hat{b}^\dagger_R(K')  \right. \nonumber \\
& \left.  + \sum_{n=3}^{\infty} \frac{n}{n!} \left[   \int_{-\infty}^{+\infty} dK' \exp \left( -\frac{\beta}{2} |K'| \right) \hat{b}^\dagger_L(K') \hat{b}^\dagger_R(K') \right]^{n-1} \right\rbrace \nonumber \\
  = & \exp \left( -\frac{\beta}{2} |K| \right) \hat{b}^\dagger_{R,L}(K)  \sum_{n=1}^{\infty} \frac{1}{(n-1)!}   \left[   \int_{-\infty}^{+\infty} dK' \exp \left( -\frac{\beta}{2} |K'| \right) \hat{b}^\dagger_L(K') \hat{b}^\dagger_R(K') \right]^{n-1} \nonumber \\
= & \exp \left( -\frac{\beta}{2} |K| \right) \hat{b}^\dagger_{R,L}(K) \hat{S}.
\end{align}
Therefore, we obtain
\begin{align} \label{from_left_to_right}
\hat{b}^\dagger_{R,L}(K)  \hat{S} | 0_L, 0_R \rangle = & \exp \left( \frac{\beta}{2} |K| \right) \left[ \hat{b}_{L,R}(K) , \hat{S} \right] | 0_L, 0_R \rangle \nonumber \\
= & \exp \left( \frac{\beta}{2} |K| \right) \hat{b}_{L,R}(K) \hat{S} | 0_L, 0_R \rangle,
\end{align}
which means that the $|0_M\rangle$ of Eq. (\ref{vacuum_state_H_M_in_H'_M}) is solution of Eqs. (5).

Eq. (5) can be used to obtain useful identities. For instance, it is possible to move creation $\hat{b}_R^\dagger(K)$ and annihilation $\hat{b}_R(K)$ operators acting from the left of $\hat{\rho}_0$ to its right and the other way round using the following identity and its adjoin
\begin{equation}\label{ordering_Rindler}
\hat{b}^\dagger_R(K) \hat{\rho}_0 = e^{\beta |K|}  \hat{\rho}_0 \hat{b}^\dagger_R(K).
\end{equation}
Eq. (\ref{ordering_Rindler}) is a result of Eq. (5):
\begin{align}
\hat{b}^\dagger_R(K) \hat{\rho}_0 = & \text{Tr}_L \left[ \hat{b}^\dagger_R(K) |0_M\rangle \langle 0_M| \right] \nonumber \\
= & \exp \left( \frac{1}{2} \beta |K| \right) \text{Tr}_L \left[ \hat{b}_L(K) |0_M\rangle \langle 0_M| \right] \nonumber \\
= & \exp \left( \frac{1}{2} \beta |K| \right) \text{Tr}_L \left[ |0_M\rangle \langle 0_M| \hat{b}_L(K) \right]\nonumber \\
= & e^{\beta |K|} \text{Tr}_L \left[ |0_M\rangle \langle 0_M| \hat{b}_R^\dagger(K) \right]\nonumber \\
= & e^{\beta |K|}  \hat{\rho}_0 \hat{b}^\dagger_R(K).
\end{align}

Eq. (\ref{ordering_Rindler}) gives also the following identity
\begin{align}\label{ordering_Rindler_2}
\hat{b}^\dagger_R(K)\hat{\rho}_0 \hat{b}_R(K') = & e^{\beta |K'|} \hat{b}^\dagger_R(K) \hat{b}_R(K') \hat{\rho}_0 \nonumber \\
= & e^{\beta |K'|} \left[ \hat{b}^\dagger_R(K), \hat{b}_R(K') \right] \hat{\rho}_0 + e^{\beta |K'|} \hat{b}_R(K') \hat{b}^\dagger_R(K) \hat{\rho}_0 \nonumber \\
= & - e^{\beta |K'|} \delta(K-K') \hat{\rho}_0 + e^{\beta (|K|+|K'|)} \hat{b}_R(K') \hat{\rho}_0 \hat{b}^\dagger_R(K),
\end{align}
which will be used in Sec. \ref{Characteristic_function}.

\section{A proof for Eq. (3)}

The main aim of this section is to provide a proof of Eq. (3) through some identities obtained in the previous sections.

Eqs. (5, \ref{Bogolyubov_transformation}, \ref{alpha_to_beta}) can be used in order to derive the following identity
\begin{align} \label{momentum_state}
\hat{a}^\dagger (k) |0_M\rangle = &  \int_{-\infty}^{+\infty}dK \left[\alpha^*(k,K)\hat{b}^\dagger_L(K) - \beta(k,K) \hat{b}_L(K) + \alpha(k,K) \hat{b}^\dagger_R(K) - \beta^*(k,K) \hat{b}_R(K) \right] |0_M\rangle \nonumber \\
= & \int_{-\infty}^{+\infty}dK \left[ \exp \left( \frac{\beta}{2} |K| \right) \alpha^*(k,K)\hat{b}_R(K) - \exp \left(-\frac{\beta}{2} |K| \right) \beta(k,K) \hat{b}^\dagger_R(K) \right. \nonumber\\
& \left. + \exp \left( \frac{\beta}{2} |K| \right)  \beta(k,K) \hat{b}^\dagger_R(K) - \exp \left(-\frac{\beta}{2} |K| \right)  \alpha^*(k,K) \hat{b}_R(K) \right] |0_M\rangle \nonumber\\
= & \int_{-\infty}^{+\infty}dK \left[ 2 \sinh \left( \frac{\beta}{2} |K| \right) \alpha^*(k,K)\hat{b}_R(K) + 2 \sinh \left( \frac{\beta}{2} |K| \right)  \beta(k,K) \hat{b}^\dagger_R(K) \right] |0_M\rangle \nonumber\\
= & \int_{-\infty}^{+\infty}dK 2 \sinh \left( \frac{\beta}{2} |K| \right) \left[ \alpha^*(k,K)\hat{b}_R(K) + \beta(k,K) \hat{b}^\dagger_R(K) \right] |0_M\rangle \nonumber\\
= &\int_{-\infty}^{+\infty}dK 2 \sinh \left( \frac{\beta}{2} |K| \right) \theta(kK) \sqrt{\frac{K}{k}} \left[ F(-k,-K)\hat{b}_R(K) + F(-k,K) \hat{b}^\dagger_R(K) \right] |0_M\rangle 
\end{align}
Eqs. (2, \ref{momentum_state}) result in
\begin{equation} \label{state}
|\psi\rangle = \int_{-\infty}^{+\infty} dK \left[ \tilde{\psi}_-(K) \hat{b}_R(K) + \tilde{\psi}_+(K) \hat{b}^\dagger_R(K) \right] |0_M\rangle ,
\end{equation}
with
\begin{equation} \label{psi_tilde} 
\tilde{\psi}_\pm(K) = 2 \sinh \left( \frac{\beta}{2} |K| \right) \int_{-\infty}^{+\infty} dk \int_{-\infty}^{+\infty} dx \frac{\psi(x)}{\sqrt{2\pi}} e^{-ikx} \theta(kK) \sqrt{\frac{K}{k}}  F(-k, \pm K).
\end{equation}

It is straightforward to obtain Eq. (3) from Eq. \ref{state} if we prove that the functions $\tilde{\psi}_\pm(K)$ defined in Eq. (\ref{psi_tilde}) are identical to Eq. (4a). This can be seen in the following way
\begin{align} 
\tilde{\psi}_\pm(K) = & 2 \sinh \left( \frac{\beta}{2} |K| \right) \int_{-\infty}^{+\infty} dk \int_{-\infty}^{+\infty} dx \frac{\psi(x)}{\sqrt{2 \pi}} e^{-ikx} \theta(kK) \sqrt{\frac{K}{k}} \frac{1}{2 \pi a} \Gamma \left(\mp \frac{i K}{a} \right) \exp \left(\pm i \frac{K}{a} \ln \frac{|k|}{a} \mp \text{sign}(k) \frac{\beta}{4} K \right) \nonumber \\
 = & 2 \sqrt{|K|} \sinh \left( \frac{\beta}{2} |K| \right) \Gamma \left(\mp \frac{i K}{a} \right) \int_{-\infty}^{+\infty} dx \frac{\psi(x)}{\sqrt{2 \pi}} \int_{-\infty}^{+\infty} dk \frac{e^{-i\text{sign}(k)|k|x}}{2 \pi a} \frac{\theta(kK)}{ \sqrt{|k|}} \exp \left(\pm i \frac{K}{a} \ln \frac{|k|}{a} \mp \text{sign}(k) \frac{\beta}{4} K \right) \nonumber \\
  = & 2 \sqrt{|K|} \sinh \left( \frac{\beta}{2} |K| \right) \Gamma \left(\mp \frac{i K}{a} \right) \int_{-\infty}^{+\infty} dx \frac{\psi(x)}{\sqrt{2 \pi}} \int_{0}^{+\infty} d|k| \frac{e^{-i\text{sign}(K)|k|x}}{2 \pi a \sqrt{|k|}}   \exp \left(\pm i \frac{K}{a} \ln \frac{|k|}{a} \mp \text{sign}(K) \frac{\beta}{4} K \right) \nonumber \\
 = & 2 \sqrt{|K|} \sinh \left( \frac{\beta}{2} |K| \right) \Gamma \left(\mp \frac{i K}{a} \right) \int_{-\infty}^{+\infty} dx \frac{\psi(x)}{\sqrt{2 \pi}} \int_{-\infty}^{+\infty} d\Xi\frac{\sqrt{a}}{2 \pi} \exp \left( \frac{a}{2} \Xi -i \text{sign}(K) a e^{a \Xi} x \pm i K \Xi \mp \frac{\beta}{4} |K| \right) \nonumber \\
 = & 2 \sqrt{a|K|} \sinh \left( \frac{\beta}{2} |K| \right) \Gamma \left(\mp \frac{i K}{a} \right) \int_{-\infty}^{+\infty} dx \frac{\psi(x)}{\sqrt{2 \pi}} F \left( - \text{sign}(K) a^2 x,\mp K + i \frac{a}{2} \right) \exp \left(\mp \frac{\beta}{4} |K| \right) \nonumber \\
 = & 2 \sqrt{a|K|} \sinh \left( \frac{\beta}{2} |K| \right) \Gamma \left(\mp \frac{i K}{a} \right) \int_{-\infty}^{+\infty} dx \frac{\psi(x)}{\sqrt{2 \pi}} \frac{1}{2 \pi a} \Gamma \left(\frac{1}{2} \pm \frac{i K}{a} \right) \exp \left( \left(\mp i \frac{K}{a} - \frac{1}{2} \right) \ln |a x| \pm \text{sign} ( x K) \frac{\beta}{4} K \right. \nonumber \\
& \left. - i \text{sign}(x K) \frac{\pi}{4} \mp \frac{\beta}{4} |K|\right) \nonumber \\
 = & \frac{\sqrt{|K|}}{\pi \sqrt{a}} \sinh \left( \frac{\beta}{2} |K| \right) \Gamma \left(\mp \frac{i K}{a} \right) \Gamma \left(\frac{1}{2} \pm \frac{i K}{a}\right)  \int_{-\infty}^{+\infty} dx \frac{\psi(x)}{\sqrt{2 \pi}} \exp \left( \left(\mp i \frac{K}{a} - \frac{1}{2} \right) \ln |a x| \pm \text{sign} ( x) \frac{\beta}{4} |K| \right. \nonumber \\
& \left. - i \text{sign}(x K) \frac{\pi}{4} \mp \frac{\beta}{4} |K|\right) \nonumber \\
 = & \frac{\sqrt{|K|}}{\pi \sqrt{a}} \sinh \left( \frac{\beta}{2} |K| \right) \Gamma \left(\mp \frac{i K}{a} \right) \Gamma \left(\frac{1}{2} \pm \frac{i K}{a}\right) \left[ \int_{0}^{+\infty} dx \frac{\psi(x)}{\sqrt{2 \pi}} \exp \left( \left(\mp i \frac{K}{a} - \frac{1}{2} \right) \ln |a x| - i \text{sign}(K) \frac{\pi}{4}\right)  \right. \nonumber \\
& \left. + \int_{-\infty}^{0} dx \frac{\psi(x)}{\sqrt{2 \pi}} \exp \left( \left(\mp i \frac{K}{a} - \frac{1}{2} \right) \ln |a x| + i \text{sign}(K) \frac{\pi}{4} \mp \frac{\beta}{2} |K|\right) \right] \nonumber \\
 = & 2 \sinh \left( \frac{\beta}{2} |K| \right) \exp \left( \mp \frac{\beta}{2} |K|\right) \left\lbrace \left[ \theta(\pm 1) + \tilde{f}_{R \pm}\left( \mp \frac{K}{a} \right) \right] \int_{-\infty}^{+\infty} dX \frac{\psi(x_R(X))}{\sqrt{2 \pi}} \exp \left( \left(\mp i K + \frac{a}{2} \right) X \right)  \right. \nonumber \\
& \left. + \tilde{f}_{L \pm}\left( \mp \frac{K}{a} \right) \int_{-\infty}^{+\infty} dX \frac{\psi(x_L(-X))}{\sqrt{2 \pi}} \exp \left( \left(\mp i K + \frac{a}{2} \right) X \right) \right\rbrace \nonumber \\
 = & \frac{e^{-\theta(\pm 1)\beta |K|}}{n_0(K)}   \left\lbrace \left[ \theta(\pm 1) + \tilde{f}_{R \pm}\left( \mp \frac{K}{a} \right) \right] \int_{-\infty}^{+\infty} dX \psi_R(X) \frac{e^{ \mp i K X }}{\sqrt{2 \pi}}   + \tilde{f}_{L \pm}\left( \mp \frac{K}{a} \right) \int_{-\infty}^{+\infty} dX \psi_L(-X) \frac{e^{\mp i K X}}{\sqrt{2 \pi}} \right\rbrace \nonumber \\
 = & \frac{e^{-\theta(\pm 1)\beta |K|}}{n_0(K)} \int_{-\infty}^{+\infty} dX  \frac{e^{\mp i K X}}{\sqrt{2 \pi}}  \left\lbrace \psi_R(X) \left[ \theta(\pm 1) + \tilde{f}_{R \pm}\left( \mp \frac{K}{a} \right) \right] + \psi_L(-X) \tilde{f}_{L \pm}\left( \mp \frac{K}{a} \right) \right\rbrace.
\end{align}

\section{Normalization conditions for single-particle states}

In this section we provide equivalent ways to express the normalization condition $\langle \psi | \psi \rangle = 1$ for different definitions of wave-functions. Moreover, we discuss the case of $\bar{\psi}(x)$ as the wave-function which transforms as a scalar field under Lorentz transformations. We show the explicit normalization condition for $\bar{\psi}(x)$ and a proof for the fact that left-wedge values of $\bar{\psi}(x)$ appear in $\Delta n_R(X)$.

The wave-function $\psi(x)$ is normalized in the following way.
\begin{equation} \label{normalization_x}
\int_{-\infty}^{+\infty} dx \left| \psi(x)\right|^2 = 1.
\end{equation}

Equivalently, the normalization condition $\langle \psi | \psi \rangle = 1$ can be imposed on the wave-function in the momentum space defined as
\begin{equation}
\tilde{\psi}(k) = \int_{-\infty}^{+\infty} dx \frac{e^{- i k x}}{\sqrt{2 \pi}} \psi(x).
\end{equation}
The normalization condition for $\tilde{\psi}(k)$ writes
\begin{equation} \label{normalization_k}
\int_{-\infty}^{+\infty} dk \left|\tilde{\psi}(k)\right|^2 = 1.
\end{equation}

Eq. (\ref{normalization_x}) can be written in terms of $\psi_R (X)$ and $\psi_L(X)$
\begin{equation} \label{normalization_x_2}
P_L + P_R = 1,
\end{equation}
where
\begin{equation}
P_{L,R} = \int_{-\infty}^{+\infty} dX \left| \psi_{L,R}(X)\right|^2
\end{equation}
are associated to the probability of finding the particle in the left and in the right wedge.

Eq. (\ref{normalization_k}) can be written in terms of $\tilde{\psi}_\pm(K)$
\begin{equation} \label{normalization_2}
\int_{-\infty}^{+\infty} dK n_0(K) \left[ \left| \tilde{\psi}_+(K) \exp \left( \frac{\beta}{2}|K| \right) \right|^2 + \left|\tilde{\psi}_-(K)\right|^2 \right] = 1.
\end{equation}
Eq. (\ref{normalization_2}) can be obtained from Eqs. (\ref{identity_K}, \ref{psi_tilde}) and from the following identity
\begin{align} \label{psi_tilde_3}
\tilde{\psi}_\pm(K) \exp \left( \pm \frac{\beta}{2} |K| \right)  = & 2 \sinh \left( \frac{\beta}{2} |K| \right) \int_{-\infty}^{+\infty} dk \tilde{\psi}(k) \theta(kK) \sqrt{\frac{K}{k}} F(k, \pm K),
\end{align}
which in turn can be directly obtained from Eqs. (\ref{F}, \ref{psi_tilde}). The proof for Eq. (\ref{normalization_2}) is the following
\begin{align}
& \int_{-\infty}^{+\infty} dK n_0(K) \left[ \left| \tilde{\psi}_+(K) \exp \left( \frac{\beta}{2}|K| \right) \right|^2 + \left|\tilde{\psi}_-(K)\right|^2 \right] \nonumber \\
 = & \int_{-\infty}^{+\infty} dK \exp \left( -\frac{\beta}{2}|K| \right) \left[ 2 \sinh\left( \frac{\beta}{2}|K| \right) \right]^{-1} \left[ \left| \tilde{\psi}_+(K) \exp \left( \frac{\beta}{2}|K| \right) \right|^2 + \left|\tilde{\psi}_-(K)\right|^2  \right] \nonumber \\
= & \int_{-\infty}^{+\infty} dK \left[ 2 \sinh\left( \frac{\beta}{2}|K| \right) \right]^{-1} \left[  \tilde{\psi}_+(K) \exp \left( \frac{\beta}{2}|K| \right) \tilde{\psi}^*_+(K) + \tilde{\psi}_-(K) \exp \left( -\frac{\beta}{2}|K| \right) \tilde{\psi}^*_-(K)  \right]\nonumber \\
= & \int_{-\infty}^{+\infty} dK 2 \sinh \left( \frac{\beta}{2} |K| \right) |K|  \int_{-\infty}^{+\infty} dk \frac{\tilde{\psi}(k)}{\sqrt{|k|}} \theta(kK) \int_{-\infty}^{+\infty} dk' \frac{\tilde{\psi}^*(k')}{\sqrt{|k'|}} \theta(k'K) [F(k, K) F(k', - K) + F(k, - K) F(k', K)]\nonumber \\
= & \int_{-\infty}^{+\infty} dk \tilde{\psi}(k) \int_{-\infty}^{+\infty} dk' \tilde{\psi}^*(k') \int_{-\infty}^{+\infty} dK 2 \sinh \left( \frac{\beta}{2} |K| \right) |K|  \frac{\theta(kK) \theta(k'K)}{\sqrt{|k k'|}} [F(k, K) F(k', - K) + F(k, - K) F(k', K)]\nonumber \\
= & \int_{-\infty}^{+\infty} dk \tilde{\psi}(k) \int_{-\infty}^{+\infty} dk' \tilde{\psi}^*(k') \delta(k-k') \nonumber \\
= & \int_{-\infty}^{+\infty} dk \left| \tilde{\psi}(k) \right|^2 \nonumber \\
= & 1.
\end{align}

$\psi(x)$ does not behave as a scalar field under Lorentz transformations. Nevertheless, we can use the definition of a Lorentz-invariant wave-function $\bar{\psi}(x)$ through the following expression
\begin{equation} \label{one_state_invariant}
|\psi\rangle = \int_{-\infty}^{+\infty} dx \bar{\psi}(x) \hat{\phi} (0,x) |0_M\rangle.
\end{equation}
The downside is that the normalization condition for $\bar{\psi}(x)$ is not provided with the interpretation of $\left|\bar{\psi}(x)\right|^2$ as the probability density of finding a particle at position $x$ \cite{Localized}.

A possible way to switch between $\tilde{\psi}(k)$ and $\bar{\psi}(x)$ can be obtained from Eqs. (2,  \ref{one_state_invariant}). Indeed, Eq. (2) is equivalent to
\begin{align} \label{one_state_alt}
|\psi\rangle = & \int_{-\infty}^{+\infty} dx \psi(x) \hat{\tilde{a}}^\dagger (x) |0_M\rangle \nonumber\\
 = & \int_{-\infty}^{+\infty} dx \psi(x) \int_{-\infty}^{+\infty} dk  \frac{e^{-i k x}}{\sqrt{2 \pi}} \hat{a}^\dagger (k) |0_M\rangle \nonumber\\
  = & \int_{-\infty}^{+\infty} dk \tilde{\psi}(k) \hat{a}^\dagger (k) |0_M\rangle,
\end{align}
while Eq. (\ref{one_state_invariant})
\begin{align} \label{one_state_invariant_alt}
|\psi\rangle = & \int_{-\infty}^{+\infty} dx \bar{\psi}(x) \hat{\phi} (0,x) |0_M\rangle \nonumber\\
= & \int_{-\infty}^{+\infty} dx \bar{\psi}(x) \int_{-\infty}^{+\infty} \frac{dk}{\sqrt{2\pi|k|}} \left[ e^{ikx} \hat{a}(k) + e^{-ikx}\hat{a}^\dagger (k) \right] |0_M\rangle \nonumber\\
= & \int_{-\infty}^{+\infty} dx \bar{\psi}(x) \int_{-\infty}^{+\infty} \frac{dk}{\sqrt{2\pi|k|}}  e^{-ikx}\hat{a}^\dagger (k) |0_M\rangle \nonumber\\
  = & \int_{-\infty}^{+\infty} dk \int_{-\infty}^{+\infty} dx  \frac{e^{-ikx}}{\sqrt{2\pi|k|}} \bar{\psi}(x) \hat{a}^\dagger (k) |0_M\rangle.
\end{align}
By comparing Eq.(\ref{one_state_alt}) with Eq. (\ref{one_state_invariant_alt}), we obtain the following identity relating $\tilde{\psi}(k)$ and $\bar{\psi}(x)$:
\begin{equation}\label{one_state_invariant_2}
 \tilde{\psi}(k) = \int_{-\infty}^{+\infty} dx  \frac{e^{-ikx}}{\sqrt{2\pi|k|}} \bar{\psi}(x).
\end{equation}
Eqs. (\ref{normalization_k}, \ref{one_state_invariant_2}) result in the following normalization condition for $\bar{\psi}(x)$
\begin{equation}\label{one_state_invariant_normalization}
\int_{-\infty}^{+\infty} dk \left| \int_{-\infty}^{+\infty} dx  \frac{e^{-ikx}}{\sqrt{2\pi|k|}} \bar{\psi}(x) \right|^2 = 1,
\end{equation}
which makes Gaussian functions non-normalizable. Indeed a necessary condition for Eq. (\ref{one_state_invariant_normalization}) is that the absolute value of the Fourier transform of $\bar{\psi}(x)$ is negligible at the origin, which is an impossible condition for Gaussian functions.

The fact that left-wedge values of $\bar{\psi}(x)$ appear in $\Delta n_R(X)$ -- as we have stated in the main paper -- can be proven in the following way:
\begin{align}\label{state_invariant}
| \psi \rangle = & \int_{-\infty}^{+\infty} dx \bar{\psi}(x) \hat{\phi}(x) | 0_M \rangle \nonumber\\
 = & \int_{-\infty}^{0} dx \bar{\psi}(x) \hat{\phi}(x) | 0_M \rangle + \int_{0}^{+\infty} dx \bar{\psi}(x) \hat{\phi}(x) | 0_M \rangle \nonumber\\
    = & \int_{-\infty}^{0} dx \bar{\psi}(x) \hat{\Phi}_L(0,X_L(x)) | 0_M \rangle + \int_{0}^{+\infty} dx  \bar{\psi}(x) \hat{\Phi}_R(0,X_R(x)) | 0_M \rangle \nonumber\\
    = & \int_{-\infty}^{0} dx \bar{\psi}(x)  \int_{-\infty}^{+\infty} \frac{dK}{\sqrt{2\pi|K|}} \left[ e^{iKX_L(x)} \hat{b}_L(K)  + e^{-iKX_L(x)} \hat{b}_L^\dagger(K) \right] | 0_M \rangle \nonumber\\
    & + \int_{0}^{+\infty} dx  \bar{\psi}(x)  \int_{-\infty}^{+\infty} \frac{dK}{\sqrt{2\pi|K|}} \left[ e^{iKX_R(x)} \hat{b}_R(K)  + e^{-iKX_R(x)} \hat{b}_R^\dagger(K) \right] | 0_M \rangle \nonumber\\
    = & \int_{0}^{+\infty} dx  \int_{-\infty}^{+\infty} \frac{dK}{\sqrt{2\pi|K|}} \left\lbrace \bar{\psi}(-x) \left[ e^{-iKX_R(x)} \hat{b}_L(K)  + e^{iKX_R(x)} \hat{b}_L^\dagger(K) \right] \right. \nonumber \\
  &  \left. +\bar{\psi}(x) \left[ e^{iKX_R(x)} \hat{b}_R(K)  + e^{-iKX_R(x)} \hat{b}_R^\dagger(K) \right]  \right\rbrace | 0_M \rangle \nonumber\\
    = & \int_{0}^{+\infty} dx  \int_{-\infty}^{+\infty} \frac{dK}{\sqrt{2\pi|K|}} e^{-iKX_R(x)} \left\lbrace \bar{\psi}(-x) \left[ \hat{b}_L(K)  + \hat{b}_L^\dagger(-K) \right]  +\bar{\psi}(x) \left[ \hat{b}_R(-K)  +  \hat{b}_R^\dagger(K) \right]  \right\rbrace | 0_M \rangle \nonumber\\
    = & \int_{0}^{+\infty} dx  \int_{-\infty}^{+\infty} \frac{dK}{\sqrt{2\pi|K|}} e^{-iKX_R(x)} \left\lbrace \bar{\psi}(-x) \left[ \exp \left( - \frac{\beta}{2}|K| \right) \hat{b}^\dagger_R(K)  + \exp \left( \frac{\beta}{2}|K| \right) \hat{b}_R(-K) \right] \right. \nonumber \\
    & \left.  +\bar{\psi}(x) \left[ \hat{b}_R(-K)  +  \hat{b}_R^\dagger(K) \right]  \right\rbrace | 0_M \rangle \nonumber\\
    = & \int_{-\infty}^{+\infty} \frac{dK}{\sqrt{2\pi|K|}} \int_{0}^{+\infty} dx   e^{-iKX_R(x)} \left\lbrace \left[ \exp \left( - \frac{\beta}{2}|K| \right) \bar{\psi}(-x) +\bar{\psi}(x) \right] \hat{b}^\dagger_R(K)  \right. \nonumber \\
    & \left.  +\left[\exp \left( \frac{\beta}{2}|K| \right) \bar{\psi}(-x) +  \bar{\psi}(x) \right] \hat{b}_R(-K)  \right\rbrace | 0_M \rangle.
\end{align}
By comparing Eq. (\ref{state_invariant}) with Eq. (3), we obtain the expression for $\tilde{\psi}_\pm (K)$ in terms of $\bar{\psi}(x)$:
\begin{equation} \label{Delta_n_invariant}
\tilde{\psi}_\pm (K) = \frac{1}{\sqrt{2\pi|K|}}\int_{0}^{+\infty} dx   e^{-iKX_R(x)} \left[ \exp \left( \mp \frac{\beta}{2}|K| \right) \bar{\psi}(-x) +\bar{\psi}(x) \right].
\end{equation}
The functions $\tilde{\psi}_\pm (K)$ appear in the final expression of $\Delta n_R(X)$ through Eqs. (11, 12b). Therefore, Eq. (\ref{Delta_n_invariant}) reveals the fact that left-wedge values of $\bar{\psi}(x)$ appear in $\Delta n_R(X)$.

\section{A proof for Eq. (8)} \label{Characteristic_function}

We want to provide a proof for Eq. (8). Firstly, it is possible to notice that, thanks to Eqs. (\ref{ordering_Rindler}, \ref{ordering_Rindler_2}), Eq. (6) can be put in the following form:
\begin{align}
\hat{\rho} = & \int_{-\infty}^{+\infty} dK \int_{-\infty}^{+\infty} dK' \left[ \tilde{\psi}_-(K) \tilde{\psi}^*_-(K') \hat{b}_R(K)\hat{\rho}_0 \hat{b}^\dagger_R(K') + \tilde{\psi}_-(K) \tilde{\psi}^*_+(K') \hat{b}_R(K)\hat{\rho}_0 \hat{b}_R(K') \right. \nonumber \\
& \left. + \tilde{\psi}_+(K) \tilde{\psi}^*_-(K') \hat{b}^\dagger_R(K)\hat{\rho}_0 \hat{b}^\dagger_R(K') + \tilde{\psi}_+(K) \tilde{\psi}^*_+(K') \hat{b}^\dagger_R(K)\hat{\rho}_0 \hat{b}_R(K') \right] \nonumber \\
 = & \int_{-\infty}^{+\infty} dK \int_{-\infty}^{+\infty} dK' \left\lbrace \tilde{\psi}_-(K) \tilde{\psi}^*_-(K') \hat{b}_R(K)\hat{\rho}_0 \hat{b}^\dagger_R(K') + \tilde{\psi}_-(K) \tilde{\psi}^*_+(K') e^{\beta |K'|} \hat{b}_R(K) \hat{b}_R(K')\hat{\rho}_0 \right. \nonumber \\
& \left. + \tilde{\psi}_+(K) \tilde{\psi}^*_-(K') e^{\beta |K|} \hat{\rho}_0 \hat{b}^\dagger_R(K)\hat{b}^\dagger_R(K') + \tilde{\psi}_+(K) \tilde{\psi}^*_+(K') e^{\beta |K'|} \left[ e^{\beta|K|} \hat{b}_R(K') \hat{\rho}_0 \hat{b}^\dagger_R(K) -\delta(K-K') \hat{\rho}_0\right] \right\rbrace .
\end{align}
This last identity, together with Eqs. (9, \ref{normalization_2}) and the definition for $\chi_0[\xi,\xi^*]$ -- i.e.
\begin{equation} \label{chi_0}
\chi_0[\xi,\xi^*] = \text{Tr} \left\lbrace \hat{\rho}_0 \exp \left[ \int_{-\infty}^{+\infty} dK  \xi(K) \hat{b}^\dagger_R(K)\right] \exp \left[- \int_{-\infty}^{+\infty} dK \xi^*(K) \hat{b}_R(K) \right] \right\rbrace
\end{equation}
-- gives
\begin{align}
\chi[\xi,\xi^*]  = &\int_{-\infty}^{+\infty} dK \int_{-\infty}^{+\infty} dK'  \left\lbrace -\tilde{\psi}_-(K) \tilde{\psi}^*_-(K') \frac{\delta}{\delta \xi^*(K)} \frac{\delta}{\delta \xi(K')} + \tilde{\psi}_-(K) \tilde{\psi}^*_+(K') e^{\beta |K'|} \frac{\delta}{\delta \xi^*(K)} \frac{\delta}{\delta \xi^*(K')} \right.  \nonumber \\
& \left. + \tilde{\psi}_+(K) \tilde{\psi}^*_-(K') e^{\beta |K|} \frac{\delta}{\delta \xi(K)} \frac{\delta}{\delta \xi(K')}  + \tilde{\psi}_+(K) \tilde{\psi}^*_+(K') e^{\beta |K'|} \left[-e^{\beta |K|} \frac{\delta}{\delta \xi(K)} \frac{\delta}{\delta \xi^*(K')} -\delta(K-K') \right] \right\rbrace\chi_0[\xi,\xi^*] \nonumber \\
 = &\int_{-\infty}^{+\infty} dK \int_{-\infty}^{+\infty} dK'  \left\lbrace \tilde{\psi}_-(K) \tilde{\psi}^*_-(K') [-n_0(K) n_0(K') \xi(K)\xi^*(K') + n_0(K') \delta(K-K') ] \right. \nonumber \\
& + \tilde{\psi}_-(K) \tilde{\psi}^*_+(K') e^{\beta |K'|} n_0(K) n_0(K') \xi(K) \xi(K')+ \tilde{\psi}_+(K) \tilde{\psi}^*_-(K') e^{\beta |K|} n_0(K) n_0(K') \xi^*(K) \xi^*(K') \nonumber \\
& \left.   + \tilde{\psi}_+(K) \tilde{\psi}^*_+(K') e^{\beta |K'|} \left[-e^{\beta |K|} n_0(K) n_0(K') \xi^*(K) \xi(K') + e^{\beta |K|} n_0(K')\delta(K-K') -\delta(K-K') \right] \right\rbrace\chi_0[\xi,\xi^*] \nonumber \\
 = &\int_{-\infty}^{+\infty} dK \int_{-\infty}^{+\infty} dK'  \left\lbrace -\tilde{\psi}_-(K) \tilde{\psi}^*_-(K') n_0(K) n_0(K') \xi(K)\xi^*(K') + \tilde{\psi}_-(K) \tilde{\psi}^*_-(K') n_0(K') \delta(K-K') \right. \nonumber \\
& + \tilde{\psi}_-(K) \tilde{\psi}^*_+(K') e^{\beta |K'|} n_0(K) n_0(K') \xi(K) \xi(K')+ \tilde{\psi}_+(K) \tilde{\psi}^*_-(K') e^{\beta |K|} n_0(K) n_0(K') \xi^*(K) \xi^*(K') \nonumber \\
& \left.   + \tilde{\psi}_+(K) \tilde{\psi}^*_+(K') e^{\beta |K'|} \left[-e^{\beta |K|} n_0(K) n_0(K') \xi^*(K) \xi(K') + n_0(K')\delta(K-K')  \right] \right\rbrace\chi_0[\xi,\xi^*] \nonumber \\
 = &\int_{-\infty}^{+\infty} dK \int_{-\infty}^{+\infty} dK'  \left\lbrace n_0(K') \delta(K-K') \left[ \tilde{\psi}_+(K) \tilde{\psi}^*_+(K') e^{\beta|K|} + \tilde{\psi}_-(K) \tilde{\psi}^*_-(K')  \right] \right. \nonumber \\
 & \left. - n_0(K) n_0(K') \left[\tilde{\psi}_-(K) \xi(K) - \tilde{\psi}_+(K) e^{\beta |K|} \xi^*(K) \right]\left[\tilde{\psi}^*_-(K') \xi^*(K') - \tilde{\psi}^*_+(K') e^{\beta |K'|} \xi(K') \right] \right\rbrace \chi_0[\xi,\xi^*] \nonumber \\
 = & \left\lbrace  \int_{-\infty}^{+\infty} dK n_0(K) \left[ \left| \tilde{\psi}_+(K) \exp \left( \frac{\beta}{2}|K| \right) \right|^2 + \left|\tilde{\psi}_-(K)\right|^2 \right]\right. \nonumber \\
 & \left. - \left| \int_{-\infty}^{+\infty} dK  n_0(K) \left[\tilde{\psi}_-(K) \xi(K) - \tilde{\psi}_+(K) e^{\beta |K|} \xi^*(K) \right] \right|^2  \right\rbrace \chi_0[\xi,\xi^*]\nonumber \\
 = & \left\lbrace 1 - \left| \int_{-\infty}^{+\infty} dK  n_0(K) \left[\tilde{\psi}_-(K) \xi(K) - \tilde{\psi}_+(K) e^{\beta |K|} \xi^*(K) \right] \right|^2  \right\rbrace \chi_0[\xi,\xi^*]
\end{align}
and hence Eq. (8) holds.

\section{Probability density function}

In the present section we show how to derive Eq. (11) from Eq. (10) and Eq. (14) from Eq. (12b).

Eq. (11) can be derived from Eq. (10) through the following chain of identities
\begin{align}
\Delta n_R(X) = & \left\langle \hat{n}_R(X) \right\rangle_{\hat{\rho}} - \left\langle \hat{n}_R(X) \right\rangle_{\hat{\rho}_0} \nonumber \\
 = &  \left. \int_{-\infty}^{+\infty} dK  \frac{e^{-i K X}}{\sqrt{2\pi}} \int_{-\infty}^{+\infty} dK' \frac{e^{i K' X}}{\sqrt{2\pi}} \frac{\delta}{\delta \xi(K)} \left(-\frac{\delta}{\delta \xi^*(K')} \right) \chi_0[\xi,\xi^*]\right|_{\xi=0} - \int_{-\infty}^{+\infty} dK  \frac{e^{-i K X}}{\sqrt{2\pi}} \int_{-\infty}^{+\infty} dK' \frac{e^{i K' X}}{\sqrt{2\pi}} \nonumber \\
 & \left. \times \frac{\delta}{\delta \xi(K)} \left(-\frac{\delta}{\delta \xi^*(K')} \right) \left| \int_{-\infty}^{+\infty} dK'' n_0(K'') \left[  - \tilde{\psi}_+(K'') e^{\beta |K''|} \xi^*(K'') + \tilde{\psi}_-(K'') \xi(K'') \right] \right|^2  \chi_0[\xi,\xi^*] \right|_{\xi=0} \nonumber \\
 &  - \left\langle \hat{n}_R(X) \right\rangle_{\hat{\rho}_0} \nonumber \\
 = &\int_{-\infty}^{+\infty} dK n_0(K)\int_{-\infty}^{+\infty} dK' n_0(K') \frac{e^{-i (K-K') X}}{2\pi}\left[ \tilde{\psi}^*_+(K) \tilde{\psi}_+(K') e^{\beta (|K|+|K'|)} + \tilde{\psi}_-(K) \tilde{\psi}^*_-(K')  \right]  \nonumber \\
 = &\left| \int_{-\infty}^{+\infty} dK n_0(K)\frac{e^{-i K X}}{\sqrt{2 \pi}} \tilde{\psi}^*_+(K) e^{\beta |K|} \right|^2 + \left| \int_{-\infty}^{+\infty} dK n_0(K)\frac{e^{-i K X}}{\sqrt{2 \pi}} \tilde{\psi}_-(K) \right|^2 \nonumber \\
 = &\left| \int_{-\infty}^{+\infty} dK n_0(K)\frac{e^{i K X}}{\sqrt{2 \pi}} \tilde{\psi}_+(K) e^{\beta |K|} \right|^2 + \left| \int_{-\infty}^{+\infty} dK n_0(K)\frac{e^{-i K X}}{\sqrt{2 \pi}} \tilde{\psi}_-(K) \right|^2 \nonumber \\
 = & n_+(X) + n_-(X).
\end{align}

It is possible to obtain Eq. (14) from Eq. (12b) and the following chain of identities
\begin{align}
n_\pm (X) = & \left| \int_{-\infty}^{+\infty} dK n_0(K)\frac{e^{\pm i K X}}{\sqrt{2 \pi}} \tilde{\psi}_\pm(K) e^{\theta(\pm 1) \beta |K|} \right|^2 \nonumber \\
= & \left| \int_{-\infty}^{+\infty} dK n_0(K)\frac{e^{- i K X}}{\sqrt{2 \pi}} \tilde{\psi}_\pm(\mp K) e^{\theta(\pm 1) \beta |K|} \right|^2 \nonumber \\
= & \left| \int_{-\infty}^{+\infty} dK \frac{e^{ - i K X}}{\sqrt{2 \pi}} \int_{-\infty}^{+\infty} dX' \frac{e^{ i K X' }}{\sqrt{2 \pi}} \left\lbrace \psi_R(X') \left[ \theta(\pm 1) + \tilde{f}_{R \pm}\left( \frac{K}{a} \right) \right]   + \psi_L( - X') \tilde{f}_{L \pm}\left( \frac{K}{a} \right)   \right\rbrace \right|^2  \nonumber \\
= & \left| \int_{-\infty}^{+\infty} dX' \left\lbrace \psi_R(X')  \left[ \theta(\pm 1) \int_{-\infty}^{+\infty} dK \frac{e^{ i K (X'-X)}}{2 \pi} + \int_{-\infty}^{+\infty} dK \frac{e^{ i K (X'-X)}}{2 \pi} \tilde{f}_{R \pm}\left( \frac{K}{a} \right) \right] \right.\right. \nonumber \\
& \left.\left.  + \psi_L( - X')  \int_{-\infty}^{+\infty} dK \frac{e^{ i K (X'-X)}}{2 \pi} \tilde{f}_{L \pm}\left( \frac{K}{a} \right)   \right\rbrace \right|^2  \nonumber \\
= & \left| \int_{-\infty}^{+\infty} dX' \left\lbrace \psi_R(X')  \left[ \theta(\pm 1) \delta(X'-X) + a f_{R \pm}(aX'-aX) \right]  + \psi_L( - X') a f_{L \pm} (aX'-aX)   \right\rbrace \right|^2  \nonumber \\
= & \left| \theta(\pm 1) \psi_R(X) + \int_{-\infty}^{+\infty} d\xi  \psi_R \left( \frac{\xi}{a} \right)  f_{R \pm}(\xi-aX)   + \int_{-\infty}^{+\infty} d\xi  \psi_L \left( - \frac{\xi}{a} \right)  f_{L \pm} (\xi-aX)  \right|^2  \nonumber \\
= &  | \theta(\pm 1) \psi_R(X) + \psi_{R \pm}(X) + \psi_{L \pm}(X) |^2 .
\end{align}

\section{Gaussian wave-packets}

The particular family of Gaussian wave-functions that we have considered are expressed by Eq. (16) with $X$ as variable and $x_0$ and $\sigma$ as parameters. In this section we want to study the behavior of $\psi_{L,R}(X)$, $\psi_{L,R \pm}(X)$ and $\Delta n_R(X)$ with respect to different values of $x_0$ and $\sigma$.

\begin{figure}[h]
\includegraphics[]{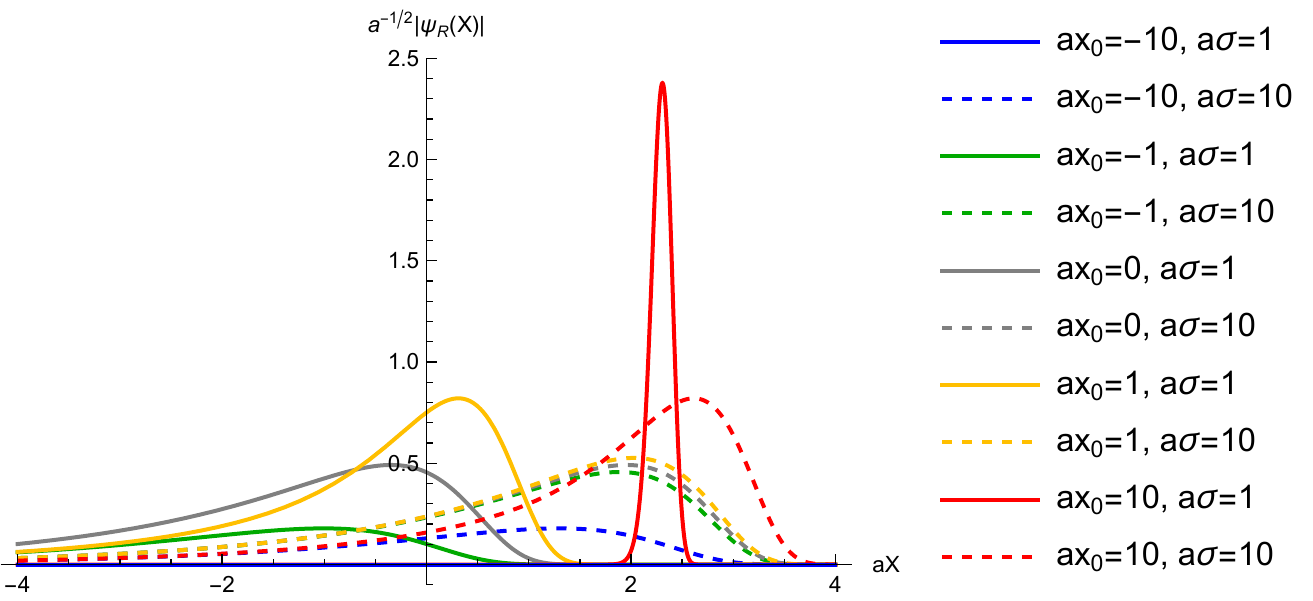}
\caption{Profile of $\psi_R(X)$ defined by Eq. (\ref{Gaussian_psi_R}) for different values of $x_0$ and $\sigma$.} \label{psi_R_figure}
\end{figure}

Notice that if we switch the sign of $x_0$ -- i.e. $x_0 \mapsto -x_0$ -- in Eq. (16), we obtain $\psi (-x)$. This results in the possibility to describe $\psi_L(X)$ through $\psi_R(X)$ with opposite $X$ and $x_0$. In other words, the transformation $X \mapsto -X$ and $x_0 \mapsto -x_0$ gives $\psi_L(X) \mapsto \psi_R(X)$ and $\psi_R(X) \mapsto \psi_L(X)$. It is, therefore, sufficient to show $\psi_R(X)$ for different parameters $x_0$ and $\sigma$ in order to describe both $\psi_R(X)$ and $\psi_L(X)$. This has been done in Fig. \ref{psi_R_figure}.

In general, it is possible to expect an almost negligible $\psi_R(X)$ for $x_0<0$ and $\sigma \ll -x_0$, since $\psi(x)$ gives most of its contribution to $\psi_L(X)$ and Eq. (\ref{normalization_x_2}) implies that $P_R \ll 1$ when $P_L \lesssim 1$. Indeed the profile for $a x_0 = -10 $ and $a \sigma = 1$ in Fig. \ref{psi_R_figure} -- marked with a blue solid line -- is not noticeable. On the other hand, a greater $\sigma$ allows $\psi(x)$ to reach the right wedge with more contribution to $\psi_R(X)$. Indeed the wave functions with $x_0 < 0$ and $\sigma \gtrsim -x_0$ are noticeable in Fig. \ref{psi_R_figure} and they are marked with dashed blue and green lines. Moreover, for greater values of $\sigma$, $P_R$ becomes larger together with the area under the curve of $\psi_R(X)$, as it is possible to notice by comparing the dashed blue and green lines with the respective solid lines. Contrary to the case of negative $x_0$, figures with larger area are the ones with $x_0>0$ and $\sigma \ll x_0$, or, in other words, wave-packets well-localized in the right wedge -- see, for instance, the solid red line in Fig. \ref{psi_R_figure}; while smaller areas are given by wave-packets that have a non-negligible probability to be found in $x<0$ -- i.e. $\sigma \gtrsim x_0$, as for the dashed red and yellow lines in Fig. \ref{psi_R_figure}.

In order to conclude the analysis of $\psi_R(X)$ for different values of $\text{sign} (x_0)$, we want to discuss the case of $x_0 = 0$. An interesting property of $\psi_R(X)$ for such case can be obtained from its explicit expression:
\begin{equation} \label{Gaussian_psi_R}
\psi_R(X) = \frac{1}{\sqrt[4]{\pi}\sqrt{\sigma}} \exp \left( \frac{a X}{2} - \frac{\left( e^{a X}-a x_0 \right)^2}{2 (a \sigma)^2}\right).
\end{equation}
Eq. (\ref{Gaussian_psi_R}) can also be put in the following form
\begin{equation} \label{Gaussian_psi_R_2}
\psi_R(X) = \frac{\sqrt{a}}{\sqrt[4]{\pi}} \exp \left( \frac{1}{2} [a X - \ln (a \sigma)]  - \frac{1}{2} \left[ e^{a X - \ln (a \sigma)}- \frac{x_0}{\sigma} \right]^2\right).
\end{equation}
In the particular case of $x_0=0$, it is easy to see from Eq. (\ref{Gaussian_psi_R_2}) that for different values of $\sigma$, the profile of $\psi_R(X)$ translates. Specifically for a transformation $\sigma \mapsto \alpha \sigma$ with $\alpha > 0$, $\psi_R(X)$ transforms in the following way: $\psi_R(X) \mapsto \psi_R(X - \ln \alpha)$. Such behavior can be observed from the gray lines in Fig. \ref{psi_R_figure}.

An equivalent expression for $\psi_R(X)$ when $x_0 \neq 0$ is the following
\begin{equation} \label{Gaussian_psi_R_3}
\psi_R(X) = \frac{\sqrt{a}}{\sqrt[4]{\pi}} \sqrt{\frac{|x_0|}{\sigma}} \exp \left( \frac{1}{2} [a X - \ln (a |x_0|)]  - \frac{1}{2} \left(\frac{|x_0|}{\sigma} \right)^2 \left[   e^{a X - \ln (a |x_0|)}- \text{sign}(x_0) \right]^2 \right).
\end{equation}
If $x_0 \neq 0$, we can use Eq. (\ref{Gaussian_psi_R_3}) to proof that any transformation $x_0 \mapsto \alpha x_0$ with $\alpha > 0$ acting on $\psi_R(X)$ is equivalent to $\sigma \mapsto \sigma / \alpha$, $a X \mapsto a X - \ln \alpha$. Equivalently, we can say that when $x_0 \neq 0$ and for fixed sign of $x_0$, the only independent variables of $\psi_R(X)$ are $X - X_R(|x_0|)$ and $|x_0|/\sigma$, as we can directly see from Eq. (\ref{Gaussian_psi_R_3}). The invariance under the transformation $x_0 \mapsto x_0/\alpha$, $\sigma \mapsto \sigma / \alpha$, $a X \mapsto a X - \ln \alpha$ for any $\alpha>0$ can be observed in Fig. \ref{psi_R_figure} through the yellow solid line with the dashed red line and the green solid line with the blue dashed line.

Thanks to the possibility to switch from $\psi_R(X)$ to $\psi_L(X)$ by performing the transformation $X \mapsto -X$ and $x_0 \mapsto -x_0$, $\psi_L(X)$ behaves similarly to $\psi_R(X)$ with respect to the property described before. Specifically, when $x_0=0$, the transformation $\sigma \mapsto \alpha \sigma$ with $\alpha > 0$ acting on $\psi_L(X)$ is equivalent to $\psi_L(X) \mapsto \psi_L(X + \ln \alpha)$. Moreover, if $x_0 \neq 0$, invariance under the transformation $x_0 \mapsto x_0/\alpha$, $\sigma \mapsto \sigma / \alpha$, $a X \mapsto a X + \ln \alpha$ for any $\alpha>0$ holds for $\psi_L(X)$. It is now straightforward to derive such properties for $\psi_{L,R\pm}(X)$ and $\Delta n_R (X)$ -- i.e. $\psi_{L,R\pm}(X)$ and $\Delta n_R (X)$ are invariant under $x_0 \mapsto x_0/\alpha$, $\sigma \mapsto \sigma / \alpha$, $X \mapsto X - \ln \alpha$ -- as we have stated in the main paper.

In summary, the behavior of $\psi_R(X)$ with respect to $|x_0|$ and for fixed $\text{sign} (x_0)$ and $\sigma/|x_0|$ corresponds to a translation with respect to $X$. The interesting limits are $a |x_0| \rightarrow 0$ and $a |x_0| \rightarrow \infty$ with finite $\sigma/|x_0|$ since for such limits the profile of $\psi_R(X)$ ``disappears'' at $a X \rightarrow - \infty$ and $a X \rightarrow + \infty$ respectively. Such behavior can be motivated in the following way. The case $a |x_0| \rightarrow 0$ and fixed $ \sigma/|x_0|$ corresponds also to the limit $a |x_0| \rightarrow 0$ and $a \sigma \rightarrow 0$, which, in turns, represents the case in which $\psi(x)$ is well-localized at the horizon. The case $a |x_0| \rightarrow \infty$ and fixed $\sigma/|x_0|$, instead, corresponds to the case in which $\psi(x)$ is localized far from the horizon. 

Since we have already explored the behavior of $\psi_R(X)$ with respect to $x_0 \neq 0$ and fixed $\sigma / |x_0|$, we are now interested in the behavior of $\psi_R(X)$ with respect to $\sigma / |x_0|$ for fixed $x_0 \neq 0$, which, in Fig. \ref{psi_R_figure}, corresponds to considering curves with the same color. In limit $\sigma / |x_0| \rightarrow \infty$ for fixed $x_0 \neq 0$, the wave-function $\psi(x)$ ``disappears'' as
\begin{equation}
\psi(x) = \mathcal{O} \left(\sqrt{\frac{|x_0|}{\sigma}}\right).
\end{equation}
In the same manner, $\psi_R(X)$ ``disappears'' since
\begin{equation}
\psi_R(X) = \mathcal{O} \left(\sqrt{\frac{|x_0|}{\sigma}}\right),
\end{equation}
as it is possible to see from Eq. (\ref{Gaussian_psi_R_3}). Moreover, the same behavior can be observed for $\psi_L(X)$, $\psi_{L,R}(X)$ and $\Delta n_R(X)$.

The limit $\sigma / |x_0| \rightarrow 0$ for fixed $x_0$ corresponds to the case of a perfectly localized Minkowski wave-function $\psi(x) = \delta(x-x_0)$. In the case of $\psi_R(X)$, Eq. (\ref{Gaussian_psi_R_3}) gives
\begin{align} \label{limit_small_sigma_x_0_R}
\lim_{\sigma/|x_0| \rightarrow 0} | \psi_R(X) |^2 = &  a e^{ a X - \ln (a |x_0|)   } \lim_{\sigma/|x_0| \rightarrow 0} \frac{|x_0|}{\sqrt{\pi} \sigma} \exp \left(  - \left(\frac{|x_0|}{\sigma} \right)^2 \left[   e^{a X - \ln (a |x_0|)}- \text{sign}(x_0) \right]^2 \right) \nonumber \\
 = & \frac{e^{a X}}{|x_0|} \delta \left( e^{a X - \ln (a |x_0|)} - \text{sign}(x_0) \right) \nonumber \\
 = &\theta(x_0) \delta \left( X - \frac{ \ln (a x_0) }{a} \right) \nonumber \\
 = & \theta(x_0) \delta(X-X_R(x_0)),
\end{align}
which can be visualized through the red and blue lines in Fig. \ref{psi_R_figure}.

Finally, the limit $\sigma / |x_0| \rightarrow 0$ for fixed $x_0$ gives the following identity
\begin{equation} \label{limit_small_sigma_x_0_a_R}
\int_{-\infty}^{+\infty} d\xi \psi_R \left( \frac{\xi}{a} \right) f(\xi)  = \theta(x_0) \sqrt{ \frac{2 \sqrt{\pi} a \sigma}{x_0}}   f(a X_R(x_0)) + o \left( \sqrt{\frac{\sigma}{|x_0|}} \right),
\end{equation}
which is valid for any function $f(\xi)$. Eq. (\ref{limit_small_sigma_x_0_a_R}) can be proven starting from the following identity
\begin{align}
 \lim_{\sigma/|x_0| \rightarrow 0} \sqrt{\frac{|x_0|}{\sigma}}\psi_R \left( \frac{\xi}{a} \right) 
 = & \lim_{\sigma/|x_0| \rightarrow 0} \frac{\sqrt{|x_0|}}{\sqrt[4]{\pi}\sigma} \exp \left( \frac{\xi}{2} - \frac{x_0^2}{2 \sigma^2} \left(\frac{ e^\xi}{a x_0} - 1\right)^2  \right)  \nonumber \\
 = & \sqrt{\frac{2\sqrt{\pi}}{|x_0|}} \exp \left( \frac{\xi}{2} \right)   \delta \left(\frac{ e^\xi}{a x_0} - 1 \right)  \nonumber \\
  = & \sqrt{2\sqrt{\pi}|x_0|} a \exp \left(-\frac{\xi}{2}  \right)  \theta(x_0) \delta (\xi -  \ln (a x_0))  \nonumber \\
= & \theta(x_0) \sqrt{ 2 \sqrt{\pi} a} \delta (\xi -  a X_R (x_0)) ,
\end{align}
which implies that
\begin{equation}
\lim_{\sigma / x_0 \rightarrow 0} \sqrt{\frac{|x_0|}{\sigma}} \int_{-\infty}^{+\infty} d\xi \psi_R \left( \frac{\xi}{a} \right) f(\xi)= \theta(x_0) \sqrt{ 2 \sqrt{\pi} a } f(a X_R(x_0))
\end{equation}
and hence Eq. (\ref{limit_small_sigma_x_0_a_R}) holds.

Thanks to the possibility of switching between $\psi_R(X)$ and $\psi_L(X)$ trough a change of sign for $X$ and $x_0$, it is possible to write Eq. (\ref{limit_small_sigma_x_0_a_R}) for $\psi_L(X)$
\begin{equation} \label{limit_small_sigma_x_0_a_L}
\int_{-\infty}^{+\infty} d\xi \psi_L \left(-\frac{\xi}{a} \right)  f(\xi)  = \theta(-x_0) \sqrt{ \frac{2 \sqrt{\pi} a \sigma}{-x_0}}  f(a X_L(x_0))+ o \left( \sqrt{\frac{\sigma}{|x_0|}} \right).
\end{equation}
Eqs. (\ref{limit_small_sigma_x_0_a_R}, \ref{limit_small_sigma_x_0_a_L}) give the following result for $\psi_{L,R \pm}(X)$
\begin{subequations}  \label{limit_small_sigma_2}
\begin{equation} \label{limit_small_sigma_L_pm}
\psi_{L \pm}(X) = \theta(-x_0) \sqrt{ \frac{2 \sqrt{\pi} a \sigma}{-x_0}}  f_{L \pm}(-a X_L(x_0) - a X) + o \left( \sqrt{\frac{\sigma}{|x_0|}} \right),
\end{equation}
\begin{equation} \label{limit_small_sigma_R_pm}
\psi_{R \pm}(X) = \theta(x_0) \sqrt{ \frac{2 \sqrt{\pi} a \sigma}{x_0}}   f_{R \pm}(a X_R(x_0) - a X)  + o \left( \sqrt{\frac{\sigma}{|x_0|}} \right),
\end{equation}
\end{subequations}
for any $X$.

The order of magnitude for vanishing values of $\psi_{L \pm}(X)$ and $\psi_{R \pm}(X)$ in the case of, respectively, $x_0>0$ and $x_0<0$ can be put in a more strict form of Eqs. (\ref{limit_small_sigma_2}) in the following way. Firstly, let us consider the following chain of inequalities
\begin{align}
\sqrt{\frac{|x_0|}{\sigma}}  \exp\left( \frac{x_0^2}{2 \sigma^2}   \right) | \psi_{R \pm}(X) | \leq &  \sqrt{\frac{|x_0|}{\sigma}}  \exp\left( \frac{x_0^2}{2 \sigma^2}   \right) \int_{-\infty}^{+\infty} d\xi \left| \psi_{R}\left(\frac{\xi}{a} \right) f_{R \pm}( \xi - aX ) \right| \nonumber \\
 \leq & \sqrt{\frac{|x_0|}{\sigma}}  \exp\left( \frac{x_0^2}{2 \sigma^2}   \right) f^{(max)}_{R \pm} \int_{-\infty}^{+\infty} d\xi \psi_{R}\left(\frac{\xi}{a} \right)\nonumber \\
= & \sqrt{\frac{|x_0|}{\sigma}}  \exp\left( \frac{x_0^2}{2 \sigma^2}   \right) \frac{f^{(max)}_{R \pm}}{\sqrt[4]{\pi}\sqrt{\sigma}} \int_{-\infty}^{+\infty} d\xi \exp \left( \frac{\xi}{2} -  \frac{1}{2}\left(\frac{ e^\xi}{a \sigma} - \frac{x_0}{\sigma}\right)^2  \right)\nonumber \\
= & \frac{f^{(max)}_{R \pm}\sqrt{|x_0|}}{\sqrt[4]{\pi}\sigma}  \exp\left( \frac{x_0^2}{2 \sigma^2}   \right) \int_{0}^{\infty} d \tilde{\xi} \sigma \sqrt{\frac{a}{|x_0|}} \exp \left( -  \frac{1}{2}\left(\frac{ \sigma \tilde{\xi}^2}{|x_0|} - \frac{x_0}{\sigma}\right)^2  \right)\nonumber \\
= & \frac{f^{(max)}_{R \pm} \sqrt{a}}{\sqrt[4]{\pi}}  \int_{0}^{\infty} d \tilde{\xi} \exp \left( -  \frac{ \sigma^2 \tilde{\xi}^4}{|x_0|^2} - \tilde{\xi}^2  \right)\nonumber \\
\leq & \frac{f^{(max)}_{R \pm} \sqrt{a}}{\sqrt[4]{\pi}}  \int_{0}^{\infty} d \tilde{\xi} \exp \left(  - \tilde{\xi}^2  \right)\nonumber \\
= & \frac{\sqrt[4]{\pi}}{2}f^{(max)}_{R \pm} \sqrt{a} ,
\end{align}
where $x_0<0$, $f^{(max)}_{R \pm}$ is the maximum value of $|f_{R \pm}( \xi)|$ and 
\begin{equation}
\tilde{\xi} = \frac{1}{\sigma} \sqrt{\frac{|x_0|}{a}e^\xi}.
\end{equation}
This leads to
\begin{equation} \label{limit_small_sigma_R_pm_2}
| \psi_{R \pm}(X) | = o \left( \sqrt{\frac{\sigma}{|x_0|}} \exp \left( -  \frac{x_0^2}{4 \sigma^2} \right) \right)
\end{equation}
when $x_0<0$. Similarly, when $x_0>0$
\begin{equation} \label{limit_small_sigma_L_pm_2}
| \psi_{L \pm}(X) | = o \left( \sqrt{\frac{\sigma}{|x_0|}} \exp \left( -  \frac{x_0^2}{4 \sigma^2} \right) \right).
\end{equation}

Eq. (\ref{limit_small_sigma_x_0_R}) results in Eq. (17a), while Eqs. (\ref{limit_small_sigma_2}, \ref{limit_small_sigma_R_pm_2}, \ref{limit_small_sigma_L_pm_2}) result in Eq. (17b).

\bibliography{bibliography} 
\bibliographystyle{ieeetr}